\documentclass{elsart}

\usepackage{epsfig}

\newcommand{\be}{\begin{equation}}
\newcommand{\ee}{\end{equation}}
\newcommand{\bea}{\begin{eqnarray}}
\newcommand{\eea}{\end{eqnarray}}

\begin{document}

\begin{frontmatter}

\title{Evidence for non-hadronic interactions of charm degrees of
freedom in heavy-ion collisions at relativistic energies}

\author[FIAS]{O.~Linnyk,\corauthref{cor1}}
\ead{linnyk@fias.uni-frankfurt.de} \corauth[cor1]{corresponding
author}
\author[FIAS]{E.~L.~Bratkovskaya,}
\author[unig]{W.~Cassing,}
\address[FIAS]{
Frankfurt Institute for Advanced Studies,%
% Ruth-Moufang-Str. 1,%
 60438 Frankfurt am Main,%
 Germany%
}%
\address[unig]{
Institut f\"ur Theoretische Physik,%
  Universit\"at Giessen,%
%  Heinrich--Buff--Ring 16,%
  35392 Giessen,%
  Germany%
}%

\date{\today}

\begin{abstract}
Within the Hadron-String Dynamics (HSD) transport approach we
study the suppression pattern of charmonia at RHIC with respect to
centrality and rapidity employing various model concepts such as
variants of the `comover absorption' model or the `charmonium
melting' scenario. We find that especially the ratio of the
forward to mid-rapidity nuclear modification factors of $J/\Psi$
($R_{AA} ^{forward} (J/\Psi) / R_{AA} ^{mid} (J/\Psi)$) cannot be
explained by the interactions with `formed' comoving mesons or by
the `color screening mechanism' alone. Only when incorporating
interactions of the $c$ or $\bar c$ quark with a pre-hadronic
medium satisfactory results are obtained. A detailed comparison to
the PHENIX data demonstrates that non-hadronic interactions are
mandatory to describe the narrowing of the $J/\Psi$ rapidity
distribution from $pp$ to central Au+Au collisions. The
 $\Psi^\prime$ to $J/\Psi$ ratio is found to be crucial in
disentangling the different charmonium absorption scenarios especially
in the RHIC energy range.  Furthermore, a comparison of the transport
calculations to the statistical model of Gorenstein and Gazdzicki  as
well as the statistical hadronization model of Andronic  {\it et al.}
 shows differences in the energy dependence as well as
centrality dependence of the $J/\Psi$ to pion ratio which may be
exploited experimentally to disentangle different concepts. We
find additionally that the collective flow of charm in the HSD
transport appears compatible with the data at SPS energies but
substantially underestimates the data at top RHIC energies such
that the large elliptic flow $v_2$ of charm seen experimentally
has to be attributed to early interactions of non-hadronic degrees
of freedom.

\end{abstract}

\begin{keyword} Relativistic heavy-ion collisions\sep
Meson production\sep Quark-gluon plasma\sep Charmed mesons \sep
Charmed quarks

PACS 25.75.-q\sep 13.60.Le\sep 12.38.Mh\sep 14.40.Lb\sep 14.65.Dw
\end{keyword}

\end{frontmatter}

%\noindent Keywords: Relativistic heavy-ion collisions; Meson
%production; Charmed mesons; Charmed quarks

%**********************************************************************

\section{Introduction}

An investigation of the formation and suppression dynamics of
$J/\Psi$, $\chi_c$ and $\Psi^\prime$ mesons opens the possibility
to address fundamental questions about the properties of the state
of matter at high temperature and density. Up to date, a
simultaneous description of the seemingly energy-independent
suppression of $J/\Psi$ together with its narrow rapidity
distribution and a strong elliptic flow $v_2$ of charmed hadrons -
as found at the Relativistic-Heavy-Ion-Collider (RHIC) - has
presented a challenge to microscopic theories. The large
discrepancies of present studies are striking in view of the
success of the hadron-string transport theories in describing
charmonium data at SPS energies. This has lead to the conjecture
that the sizeable difference between the measured yields and
transport predictions is due to a neglect of the  transition from
hadronic to partonic matter, e.g. a strongly-coupled
Quark-Gluon-Plasma (sQGP). In the present work, we report new
results on the charmonium nuclear modification factor $R_{AA}$,
rapidity distribution, the elliptic flow $v_2$ of $D$ mesons, the
ratios $\langle  J/\Psi \rangle / \langle \pi \rangle$ and
$\Psi^\prime /(J/\Psi)$ for  energies from about 20 A$\cdot$GeV -
relevant for the future
Facility-for-Antiproton-and-Ion-Research~(FAIR) - up to top RHIC
energies.

We recall that in the early stage of the nucleus-nucleus
collisions the disso\-ciation and the regeneration of $J/\Psi$ by
fundamentally different mechanisms are possible:  The $c\bar{c}$
pairs produced early in the reaction - by gluon-gluon fusion in
primary nucleon-nucleon interactions - might be completely
dissociated in the dense medium and not be formed as bound states
due to color screening. In this model scenario charmonia have to
be recreated by some mechanism to yield a finite production cross
section of $J/\Psi$ and $\Psi^\prime$. The $c\bar{c}$ pairs might
also be formed in some pre-hadronic resonance (color-dipole) state
that will further develop to the charmonium eigenstates in vacuum.
Such resonance states can be dissociated in the medium due to
interactions with other degrees of freedom but also be recreated
by the inverse reaction channels.  Independently, charmonia might
also be generated in a statistical fashion at the phase boundary
between the QGP and an interacting hadron gas such that their
abundance would appear in statistical (chemical) equilibrium with
the light and strange hadrons~\cite{PBM97,MG**2}. In the latter
model the charmonium spectra carry no information on a possible
preceeding partonic phase. Indeed, in Ref.  \cite{PBM07} a success
of the statistical hadronization model~\cite{BMS,BMS2} has been
put forward. Another alternative is the model for coalescence of
charmonium in the sQGP~\cite{Rafelski}. For further variants or
model concepts for charmonium suppression/enhancement we refer the
reader to the reviews \cite{rev1,rev2}. In this work our aim is to
shed some light on various model concepts by exploiting
relativistic microscopic transport theory.

The Hadron-String-Dynamics (HSD) approach \cite{Cass99} provides
the space-time geometry of nucleus-nucleus reactions and a rather
reliable estimate for the local energy densities achieved, since
the production of secondary particles with light and single
strange quarks/antquarks is described well from SIS to RHIC
energies~\cite{Weber}. As we will show in Section 2, the high
energy-densities reached in $Au+Au$ collisions at  RHIC clearly
indicate that a strongly interacting QGP (sQGP) has been created
for a couple of fm/c in the central overlap volume.  However, a
proper  understanding of the transport properties of the partonic
phase is still lacking. Presently the effective degrees of freedom
in the sQGP are much debated and charmonia are a unique and
promising probe that is sensitive to the properties of the early
(and so far unknown) medium.

In the present systematic study  we first test the HSD results for
charmonium production in $p+p$ and $d+Au$ reactions at RHIC
energies in comparison to the recent data.  This is crucial in
order to obtain an accurate baseline for the study of any
anomalous suppression of charmonia in nucleus-nucleus collisions
(see Sections~\ref{production} and~\ref{baryonic}).  The
interactions of $J/\Psi$'s with mesons in the late stages of the
collision (when the energy density falls below a critical value of
about 1 GeV/fm$^3$ corresponding roughly to the critical energy
density for a parton/hadron phase transition) gives a sizable
contribution to its anomalous suppression at all beam energies as
demonstrated in Refs.~\cite{Olena,Olena2,Cass01,brat03,brat04}.
Accordingly, this more obvious `hadronic' contribution has to be
incorporated when comparing possible models for QGP-induced
charmonium suppression to experimental data. On the other hand, as
known from our studies in Refs. \cite{Olena,Olena2} charmonium
interactions with the purely hadronic medium alone (which is
modeled rather precisely by HSD) are not sufficient to describe
the $J/\Psi$ suppression pattern at RHIC in detail.

Based on the microscopic HSD transport theory, we investigate in
particular the following scenarios for the anomalous absorption of
charmonia:\\ (1) the `threshold melting' mechanism;\\ (2) a
dissociation by the scattering on hadron-like correlators, {\it
i.e.} the `comover' scenario;\\ (3) additional scattering of charm
with pre-hadrons which might be considered as color neutral
precursors of hadronic states (cf. Refs.
\cite{Falter1,Falter2,HPT1,HPT2}).\\

All implemented scenarios will be described in detail in
Section~\ref{scenarios}.  In Section~\ref{results} we will
investigate in particular the effect of the interactions of charm
quarks  in the pre-hadronic medium on $R_{AA}(y)$ of $J/\Psi$ by
comparing our calculations to  RHIC data.  To complete our study,
we will provide excitation functions for the $J/\Psi$ survival
probability $S(J/\Psi)$ and the ratios $B_{\mu \mu} \sigma
(J/\Psi) / B_{\mu \mu}' \sigma (\Psi') $  in
Section~\ref{excitation}.  Furthermore, by studying the $J/\Psi$
to $\pi$ ratio as a function of the number of participating
nucleons $N_{part}$, we will test the assumption of charmonium
production by statistical hadronization as advocated in
Refs.~\cite{MG**2,PBM07,PBM} (subsection~\ref{statistical}). A
summary of results as well as a discussion of open problems will
close our study in Section 8.

%********************************************************************
\section{Energy density} \label{energydensity}

\begin{figure}
%\phantom{a}\vspace*{-1.5cm}
\centerline{\psfig{figure=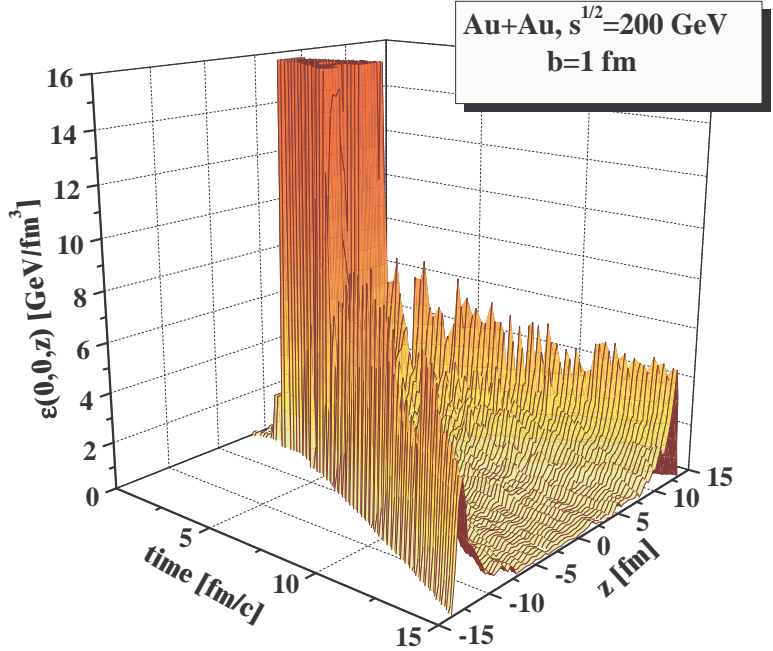,width=0.75\textwidth}}
\caption{The energy density $\varepsilon(x=0,y=0,z;t)$ from HSD
for a central Au+Au collision at $\sqrt{s}$ = 200 GeV.
The time $t$ is given in the
nucleon-nucleon center-of-mass system.} \label{3D}
\end{figure}

The HSD transport model - employed in a large variety of $\pi+A$,
$p+A$, $d+A$ and $A+A$ reactions - allows to calculate the
energy-momentum tensor $T_{\mu \nu}(x)$ for all space-time points $x$
and thus the energy density $\varepsilon(x) = T_{00}(x)$ in the local
rest frame. In order to exclude contributions to $T_{\mu \nu}$ from
noninteracting nucleons in the initial phase all nucleons without prior
interactions are discarded in the rapidity intervals $[y_{tar}-0.4,
y_{tar}+0.4]$ and $[y_{pro}-0.4, y_{pro}+0.4]$ where $y_{tar}$ and
$y_{pro}$ denote projectile and target rapidity, respectively. Note
that the initial rapidity distributions of projectile and target
nucleons are smeared out due to Fermi motion by about $\pm 0.4$. Some
comments on the choice of the grid in space-time are in order here:  In
the actual calculation (for Au+Au collisions) the initial grid has a
dimension of 1 fm $\times$ 1 fm $\times$ 1/$\gamma_{cm}$ fm, where
$\gamma_{cm}$ denotes the Lorentz $\gamma$-factor in the
nucleon-nucleon center-of-mass system. After the time of maximum
overlap $t_m$ of the nuclei the grid-size in beam direction $\Delta z_0
= 1/\gamma_{cm}$ [fm] is increased linearly in time as $\Delta z =
\Delta z_0 + a (t-t_m)$, where the parameter $a$ is chosen in a way to
keep the particle number in the local cells of volume $\Delta V(t) = \Delta x
\Delta y \Delta z(t)$ roughly constant during the
longitudinal expansion of the system. In this way local fluctuations of
the energy density $\varepsilon(x)$ due to fluctuations in the particle
number are kept low. Furthermore, the time-step is taken as $\Delta t =
0.2 \Delta z(t)$ and increases in time in analogy to $\Delta z(t)$.  This
choice provides a high resolution in space and time for the initial
phase and keeps track of the relevant dynamics throughout the entire
collision history.

The energy density $\varepsilon({\bf r};t)$ -- which is identified
with the matrix element $T_{00}({\bf r};t)$ of the energy momentum
tensor in the local rest frame at space-time $({\bf r},t)$ --
becomes very high in a central Au+Au collision at $\sqrt{s}$ = 200
GeV as shown in Fig.\ref{3D} ({\it cf.} Fig.~1 of~\cite{Olena} for
the corresponding energy density evolution in case of central
collisions at top SPS energies). Fig.~\ref{3D} shows the
space-time evolution of the energy density
$\varepsilon(x=0,y=0,z;t)$ for a Au+Au collision at 21300~AGeV or
$\sqrt{s} $ = 200 GeV. It is clearly seen that energy densities
above 16 GeV/fm$^3$ are reached in the early overlap phase of the
reaction and that $\varepsilon(x)$ drops after about 6~fm/c
(starting from contact) below 1 GeV/fm$^3$ in the center of the
grid. On the other hand the energy density in the region of the
leading particles - moving almost with the velocity of light -
stays above 1 GeV/fm$^3$ due to Lorentz time dilatation since the
time $t$ in the transport calculation is measured in the
nucleon-nucleon center-of-mass system. Note that in the local rest
frame of the leading particles the eigentime $\tau$ is roughly
given by $\tau \approx t/\gamma_{cm}$.  As seen from Fig.~1, the
energy density in the local rest frame is a rapidly changing
function of time in nucleus-nucleus collisions.  For orientation
let us recall the relevant time scales (in the cms reference
frame):

-- The $c\bar{c}$ formation time $\tau_c \approx 1/M_\perp$ is
about 0.05 fm/c for a transverse mass of 4 GeV; the transient time
for a central Au+Au collision at $\sqrt{s}$ = 200 GeV is $t_r
\approx 2 R_A/\gamma_{cm} \approx 0.13$ fm/c.  According to
standard assumptions, the $c\bar{c}$ pairs are produced in the
initial hard $NN$ collisions dominantly by gluon fusion in the
time period $t_r$.  In fact, the formation time $\tau_c$ is
significantly smaller than $t_r$, which implies the $c$ or
$\bar{c}$ quarks may interact with the impinging nucleons of the
projectile or target for times $t \leq t_r$.

-- Using the Borken estimate for the energy density and employing
the time-scale $t_r = 0.13$ fm/c, the energy density -- after the
nuclei have punched through each other -- amounts to about $5/0.13
> 30$
 GeV/fm$^3$ as quoted also in the HSD calculations in
Ref.~\cite{Olena}. Even when adding the $c \bar{c}$ formation
time, this gives an energy density $\sim 5/0.18 \approx 28$
GeV/fm$^3$.  So these numbers agree with transparent and simple
estimates (cf. Fig.~\ref{3D}) and illustrate the high initial
densities after $c\bar{c}$ production from primary interactions.

\begin{figure}
\centerline{\psfig{figure=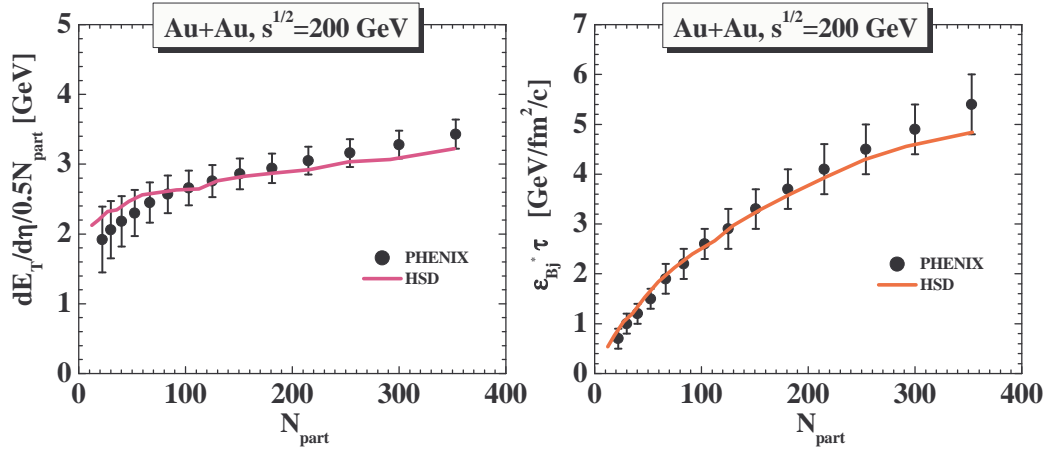,width=\textwidth}}
\caption{Left part: The energy density $E_T$ per pseudorapidity interval
$d \eta$ divided by the number of
participant pairs $ (0.5 N_{part})$ from HSD
(solid line) in comparison to the PHENIX data (dots)~\cite{Et}. Right part:
The Bjorken energy density $\varepsilon_{Bj}\cdot \tau$ from HSD
(solid line) for Au+Au collisions at $\sqrt{s}$ = 200 GeV in
comparison to the PHENIX data (dots)~\cite{Et}. } \label{ET}
\end{figure}

The energy densities quoted above are considerably different from
the Bjorken estimate \begin{equation} \label{bjorken}
\tau \cdot \epsilon_{Bj} = \frac{<E_T> {dN \over d\eta}}{\pi R_T^2},
\end{equation}
where $<E_T>$ is the average transverse energy per particle, $dN/d
\eta$ the number of particles per unit of pseudorapidity, and
$\tau$ a formation time parameter often used as $\tau = 1$ fm/c.
Furthermore, $\pi R_T^2$ denotes the overlap area for the
corresponding centrality. Is is important to point out that the
estimate (\ref{bjorken}) is only well defined for the product
$\tau \epsilon_{Bj}$! The question naturally arises, if the
transport calculations follow the corresponding experimental
constraints.

To this aim we show $dE_T/d\eta$ (divided by half the number of
participants $N_{part}$) from HSD\footnote{The open source code is
available from Ref. \cite{open}} (l.h.s.) in comparison to the
measurements by PHENIX~\cite{Et}.  Accordingly, the Bjorken energy
density $\epsilon _{Bj}$ -- multiplied by the time-scale $\tau$
(\ref{bjorken})-- from HSD is shown additionally in the r.h.s. in
comparison to the PHENIX measurements as a function of $N_{part}$.
The similarity between the calculated quantities and the
experimental data demonstrates that the space-time evolution of
the energy-momentum tensor $T_{\mu \nu}$ in HSD is sufficiently
well under control. We now may step on with the actual
investigation of the charmonium dynamics.

%********************************************************************
\section{Implementation of charmonium production} \label{production}

\begin{figure}
%\phantom{a}
%\vspace*{-0.5cm}
\centerline{\psfig{figure=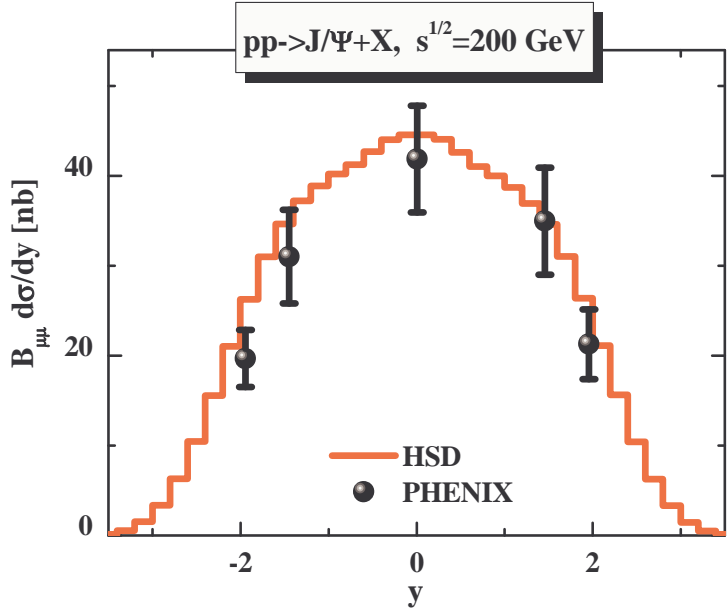,width=0.7\textwidth}}
\caption{Cross section for the differential  $J/\Psi$ production in rapidity
(times the branching ratio to di-muons $B_{\mu \mu}$) in $p p$ collisions
at $\sqrt{s}=200$~GeV. The HSD (input)
parametrization (solid line) is compared to the PHENIX data
(symbols) from Ref.~\cite{PHENIXppY}. }
\label{JPsiInpp}
\end{figure}

In order to examine the dynamics of open
charm and charmonium degrees of freedom during the formation and
expansion phase of the highly excited system created in a
relativistic nucleus-nucleus collision, one has to know the number of initially produced
particles with $c$ or $\bar{c}$ quarks, i.e. $D, \bar{D}, D^*$,
$\bar{D}^*, D_s, \bar{D}_s, D_s^*, \bar{D}_s^*,$ $J/\Psi(1S),
\Psi^\prime(2S), \chi_c(1P)$. In this work we follow the previous
studies in Refs. \cite{Cass99,Olena,Cass01,brat03,Cass00} and fit  the
total charmonium cross sections ($i = \chi_c, J/\Psi,
\Psi^\prime$) from $NN$ collisions as a function of the invariant
energy $\sqrt{s}$ by the expression \cite{Olena2}
\begin{eqnarray}
\sigma_i^{NN}(s) = f_i \ a \ \left(1 -
\frac{m_i}{\sqrt{s}}\right)^\alpha \
\left(\frac{\sqrt{s}}{m_i}\right)^\beta
\theta(\sqrt{s}-\sqrt{s_{0i}}),
 \label{fitj}
\end{eqnarray}
where $m_i$ denotes the mass of charmonium $i$ while
$\sqrt{s_{0i}}=m_i+2 m_N$ is the threshold in vacuum. The
parameters in (\ref{fitj}) have been fixed to describe the
$J/\Psi$ and $\Psi^\prime$ data up to the RHIC energy
$\sqrt{s}=200$ GeV ({\it cf.}~\cite{Olena}). We use $a=0.16$ mb,
$\alpha$ = 10, $\beta =0.775$. Note, that for the present study we
updated our parametrization (i.e. reduced the parameter $a$ by $\sim 20\%$) according
to the latest PHENIX data \cite{PHENIXppY}.

The parameters $f_i$ are fixed as
$f_{\chi_c}=0.636, \ f_{J/\Psi}=0.581,\ f_{\Psi^\prime}=0.21$ in
order to reproduce the experimental ratio $$\frac{B(\chi_{c1}\to
J/\Psi)\sigma_{\chi_{c1}}
 +B(\chi_{c2}\to J/\Psi)\sigma_{\chi_{c2}}}
 {\sigma^{exp}_{J/\Psi}}=0.344\pm 0.031$$
measured in $pp$ and $\pi N$ reactions~\cite{E705_93,WA11_82} as
well as the averaged $pp$ and $pA$ ratio
$(B_{\mu\mu}(\Psi^\prime)\sigma_{\Psi^\prime})
 / (B_{\mu\mu}(J/\Psi)\sigma_{J/\Psi})\simeq 0.0165$ ({\it cf.} the compilation
of experimental data in Ref.~\cite{NA50_03}). The experimentally
measured $J/\Psi$ cross section includes the direct $J/\Psi$
component $(\sigma_{J/\Psi})$ as well as the decays of higher
charmonium states $\chi_{c}$ and $\Psi^\prime$, {\it i.e.}
\begin{eqnarray}
\sigma^{exp}_{J/\Psi}=\sigma_{J/\Psi}+B(\chi_{c}\to
J/\Psi)\sigma_{\chi_{c}} +B(\Psi^\prime\to
J/\Psi)\sigma_{\Psi^\prime}. \ \label{xsexp}\end{eqnarray}
Note, we do not distinguish the $\chi_{c1}(1P)$ and $\chi_{c2}(1P)$
states. Instead, we use only the $\chi_{c1}(1P)$ state (which we denote
as $\chi_c$), however, with an increased  branching ratio for the decay
to $J/\Psi$ in order to include the contribution of $\chi_{c2}(1P)$,
{\it i.e.}  $B(\chi_{c}\to J/\Psi) = 0.54$. Furthermore, we adopt
$B(\Psi^\prime\to J/\Psi)=0.557$ from Ref.~\cite{PDG}.

We recall that (as in Refs.
\cite{Cass01,brat03,Geiss99,Cass97,CassKo}) the charm degrees of
freedom in the HSD approach are treated perturbatively and that
initial hard processes (such as $c\bar{c}$ or Drell-Yan production
from $NN$ collisions) are `pre-calculated' to achieve a scaling of
the inclusive cross section with the number of projectile and
target nucleons as $A_P \times A_T$ when integrating over impact
parameter. For fixed impact parameter $b$, the $c\bar{c}$ yield
then scales with the number of binary hard collisions $N_{coll}$
({\it cf.} Fig. 8 in Ref.~\cite{Cass01}).

In addition to primary hard $NN$ collisions, the open charm mesons
or charmonia may also be generated by secondary meson-baryon
($mB$) reactions. Here we include all secondary collisions of
mesons with baryons by assuming that the open charm cross section
(from Section 2 of Ref.~\cite{Cass01}) only depends on the
invariant energy $\sqrt{s}$ and not on the explicit meson or
baryon state. Furthermore, we take into account all interactions
of `formed' mesons -- after a formation time of $\tau_F$ = 0.8
fm/c (in their rest frame)~\cite{Geiss} -- with baryons or
diquarks. For the total charmonium cross sections from
meson-baryon (or $\pi N$) reactions we use the parametrization (in
line with Ref. \cite{Vogt99}):
\begin{eqnarray}
\sigma_i^{\pi N} (s) = f_i \ b \ \left(1 -
\frac{m_i}{\sqrt{s}}\right)^\gamma \label{fitpin}\end{eqnarray}
with $\gamma=7.3$ and $b=1.24$~mb, which describes the existing
experimental data at low $\sqrt{s}$ reasonably well, as seen in
Ref.~\cite{Olena}.

\begin{figure}
%\phantom{a}
%\vspace*{-0.5cm}
\centerline{\psfig{figure=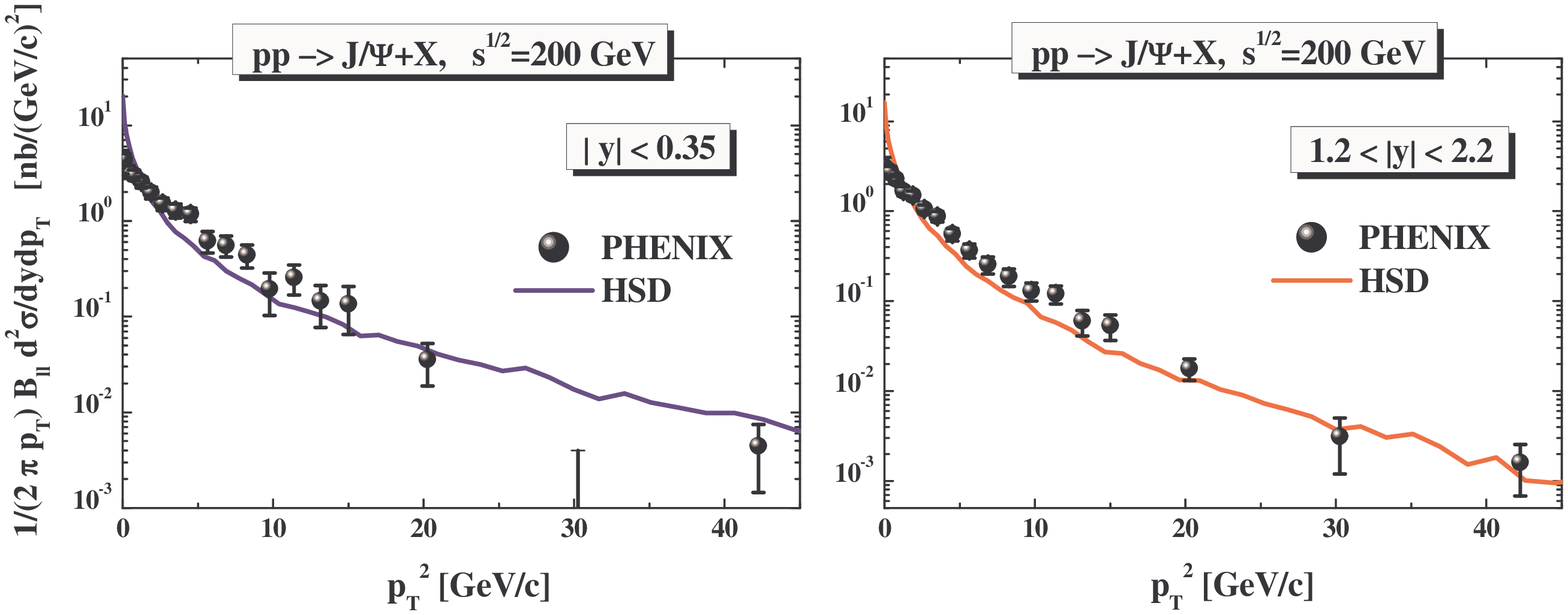,width=\textwidth}}
\caption{Differential cross section of $J/\Psi$ production in
$pp$ collisions at $\sqrt{s}=200$~GeV at mid-rapidity ($|y|<0.35$,
l.h.s.) and at forward rapidity ($1.2<|y|<2.2$, r.h.s.) {\it vs}
the transverse momentum squared $p_T^2$ as implemented in HSD
(solid line) compared to the PHENIX data from Ref.~\cite{PHENIXppY} (dots).}
\label{ppPT2mid}
\end{figure}

Apart from the total cross sections for charmonia  we also need the
differential distribution of the produced mesons in the transverse
momentum $p_T$ and the rapidity $y$ (or Feynman $x_F$) from each
individual collision. We recall that $x_F = p_z/p_z^{max} \approx 2
p_z/\sqrt{s}$ with $p_z$ denoting the longitudinal momentum. For the
differential distribution in $x_F$ from $NN$ and $\pi N$ collisions we
use the ansatz from the E672/E706 Collaboration~\cite{E672}
and for the $p_T$ distribution a power low parametrization from
Ref. \cite{brat05} which has been fixed by the STAR
data \cite{STAR04}, i.e.

\begin{equation}
\frac{dN}{dx_F dp_T} \sim (1 - |x_F|)^c \
\left(1 + {p_T\over b_{p_T}}\right)^{c_{p_T}},
\label{fit2}
\end{equation}
with $b_{p_T}=3.5$ GeV/$c$ and $c_{p_T}=-8.3$.
The exponent $c$ is given by $c= a/(1+b/\sqrt{s})$ and the parameters
$a, b$ are chosen as
$a_{NN}=16$, $b_{NN}=24.9$ GeV for $NN$ collisions and $a_{\pi N}=4.11$,
$b_{\pi N}=10.2$ GeV for $\pi N$ collisions.

The resulting rapidity distribution for $J/\Psi$ production in $pp$ collisions at
$\sqrt{s}=200$ GeV is shown in Fig.~\ref{JPsiInpp} which is in
line with the data from  Ref.~\cite{PHENIXppY} within error bars.
We also present the $p p \to
J/\Psi+X$ differential cross section in $p_T^2$ at mid-rapidity
($|y|<0.35$) and at forward rapidity (averaged in the interval
$1.2<|y|<2.2$) in Fig.~\ref{ppPT2mid}.  The HSD parametrization
is compared to the recent measurements of the corresponding
quantities by PHENIX~\cite{PHENIXppY}. The total cross sections
for $D+\bar D$ production in this study are the same as those
presented in Ref.~\cite{Olena}.

The parametrizations of the total and differential cross sections for
open charm mesons from $pN$ and $\pi N$ collisions are taken as in
Refs.~\cite{Cass01,brat03}, apart from a readjustment of the parameter
$a_{NN}$ in order to reproduce the recently measured rapidity
distribution of $J/\Psi$ in $p+p$ reactions at $\sqrt{s}=200$~GeV by
PHENIX~\cite{PHENIXppY}.

For $D, D^*, \bar{D}, \bar{D}^*$ - meson ($\pi, \eta, \rho, \omega$)
scattering we refer to the calculations from Ref.  \cite{Konew,Ko}
which predict elastic cross sections in the range of 10--20 mb
depending on the size of the form factor employed. As a guideline we
use a constant cross section of 10 mb for elastic scattering with
formed mesons and also baryons, although the latter might be even
higher for very low relative momenta. We will discuss this issue in
more detail in Section~\ref{comover}.

%***************************************************************************
\section{Baryonic (`normal') nuclear absorption} \label{baryonic}

\begin{figure}
%\phantom{a}
%\vspace*{-0.5cm}
\centerline{\psfig{figure=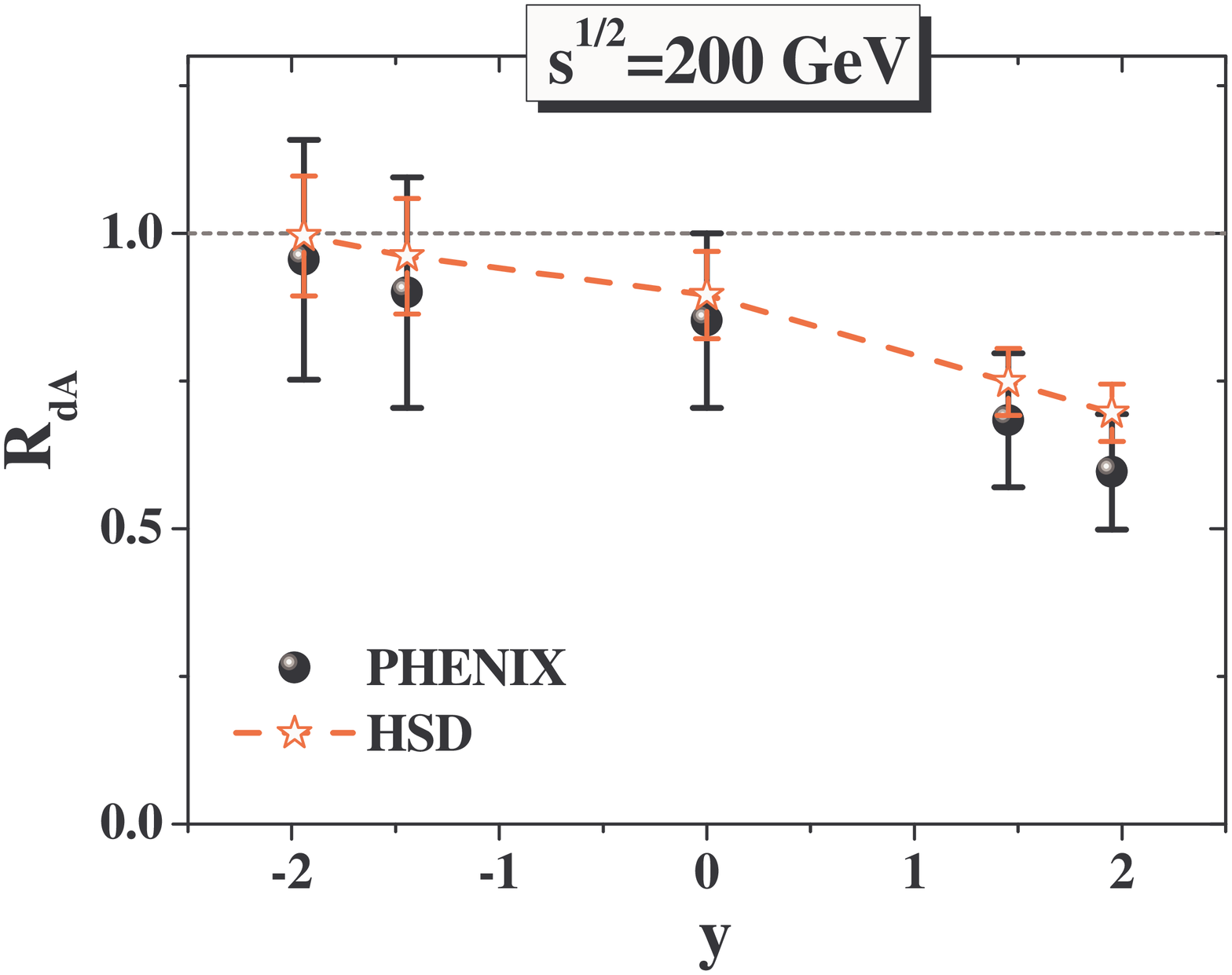,width=0.75\textwidth}}
\caption{$J/\Psi$ production cross section in $d+Au$ collisions
relative to that in $p+p$ collisions (see text for the definition
of $R_{dA}$) in HSD (red stars) as compared to the PHENIX
data~\cite{PHENIX08} (full dots).} \label{dAu}
\end{figure}

The yield of $J/\Psi$ in $p+A$ and $A+A$ reactions is modified
compared to that in $p+p$ scaled with the number of initial binary
scatterings $N_{coll}$~\cite{NA60,PHENIXNov06}. Indeed, the produced
$c\bar c$ can be dissociated or absorbed on either the residual nucleus
of the projectile or target or on light co-moving particles (usually on
mesons or, at high energy, on partons) produced in the very early
 phase.  The latter reactions are only important in nucleus-nucleus
collisions and not in $p+A$ or $d+A$ as the number of `comovers'
created in proton- or deuteron-induced processes is small. In contrast,
charmonium absorption on baryons is the leading suppression mechanism
in $d+A$ ($p+A$) scattering and is an important base-line for the study
of the absorption in the hot and dense medium created in $A+A$
reactions.

In order to study the effect of charmonium rescattering on
projectile/target nucleons, we adopt in HSD the following dissociation
cross sections of charmonia with baryons independent of the energy:
\begin{eqnarray}
&& \sigma_{c\bar{c}B} = 4.18 \ {\rm mb}; \label{sigmacB} \\
&&\sigma_{J/\Psi B} = 4.18 \ {\rm mb}; \ \sigma_{\chi_c B} = 4.18
\ {\rm mb}; \  \sigma_{\Psi^\prime B} = 7.6 \ {\rm mb}.
\nonumber\end{eqnarray}
In (\ref{sigmacB}) the cross section $\sigma_{c\bar{c}B}$ stands
for a (color dipole)  pre-resonance ($c\bar{c})$ - baryon cross
section, since the $c\bar{c}$ pair produced initially cannot be
identified with a particular charmonium due to the uncertainty
relation in energy and time. For the life-time of the
pre-resonance $c\bar{c}$ pair (in it's rest frame) a value of
$\tau_{c\bar{c}}$ = 0.3 fm/c is assumed following
Ref.~\cite{Kharz}. This time scale corresponds to the mass
difference of the $\Psi^\prime$ and $J/\Psi$.

The values for the cross sections $\sigma_{J/\Psi N},
\sigma_{c\bar c N}$ at RHIC energies are currently debated in the
literature. On one side, all the data on the $J/\Psi$
production in $p+A$ at energies $\sqrt{s} \le 40$~GeV were
found to be consistent with an energy-independent cross section
 of the order of $4-7$ {\rm
mb}~\cite{NA60,NA50pA,Borges,csPsiN1,csPsiN2}. On the other hand,
the corresponding cross sections at the much higher energy of
$\sqrt{s}=200$~GeV, {\it e.g.} at RHIC, are expected to be
smaller~\cite{dA}, since part of the suppression might be
attributed to other (initial-state) cold-matter effects, such as
gluon shadowing~\cite{Bravina,CapellaGluon,Vogt}, radiative gluon
energy loss in the initial state or multiple gluon rescattering.
We recall that `shadowing' is a depletion of low-momentum partons
in a nucleon embedded in a nucleus compared to the population in a
free nucleon, which leads to a lowering in the charmonium
production cross section. The reasons for depletion, though, are
numerous, and models of shadowing vary accordingly. There is,
therefore, a considerable (about a factor of 3) uncertainty in the
amount of shadowing predicted at
RHIC~\cite{Bravina,CapellaGluon,Vogt,Kopeliovich,Capellanew}. In
the analysis of the $d+Au$ data at $\sqrt{s}=200$~GeV, in which
the maximum estimate for the effect of the shadowing was
made~\cite{dA,Vogt}, the additional absorption on baryons allowed
by the data was found to lead to $\sigma_{J/\Psi N}=1-3$~mb or
higher, if some contribution of anti-shadowing is present. The
authors of~\cite{Vogt} advocate $\sigma_{J/\Psi N}=3$~mb in order
to preserve the agreement with the data of the Fermilab experiment
E866. The PHENIX Collaboration \cite{PHENIX08} finds a breakup
cross section of $2.8^{+1.7}_{-1.4}$ mb (using EKS shadowing)
which still overlaps with the CERN value of 4.18 mb (though with
large error bars). However, the theoretical uncertainty is still
large, since in the works above only an approximate model for
baryonic absorption was applied and not a microscopic transport
approach that e.g. also includes secondary production channels of
charm pairs as described in Section 3.

Within HSD we have found the baryoninc absorption cross sections
(\ref{sigmacB}) to agree with the data at SPS
energies~\cite{Olena}. In Fig.~\ref{dAu} we compare the HSD result
(employing the same cross sections (\ref{sigmacB}) for baryonic
absorption and neglecting shadowing) for the $J/\Psi$ production
in $d+Au$ collisions at $\sqrt{s}=200$~GeV to the inclusive PHENIX
data~\cite{PHENIX08}. The quantity plotted is the nuclear
modification factor defined as
\be
R_{dA} \equiv \frac{d N^{d Au} _{J/\Psi} / d y }{ \langle N_{coll}
\rangle \cdot d N^{p p} _{J/\Psi} / d y }, \label{FigRdA} \ee
where $d N^{d Au} _{J/\Psi} / d y$ is the $J/\Psi$ invariant yield
in $d+A$ collisions, $d N^{pp} _{J/\Psi} / d y$ is the $J/\Psi$
invariant yield in $p+p$ collisions; $\langle N_{coll} \rangle$ is
the average number of binary collisions for the same rapidity bin.
In our analysis we have used $\langle N_{coll} \rangle=7.6\pm 0.3$
according to the PHENIX estimate~\cite{PHENIX08}.

\begin{figure}
%\phantom{a}
%\vspace*{-0.5cm}
\centerline{\psfig{figure=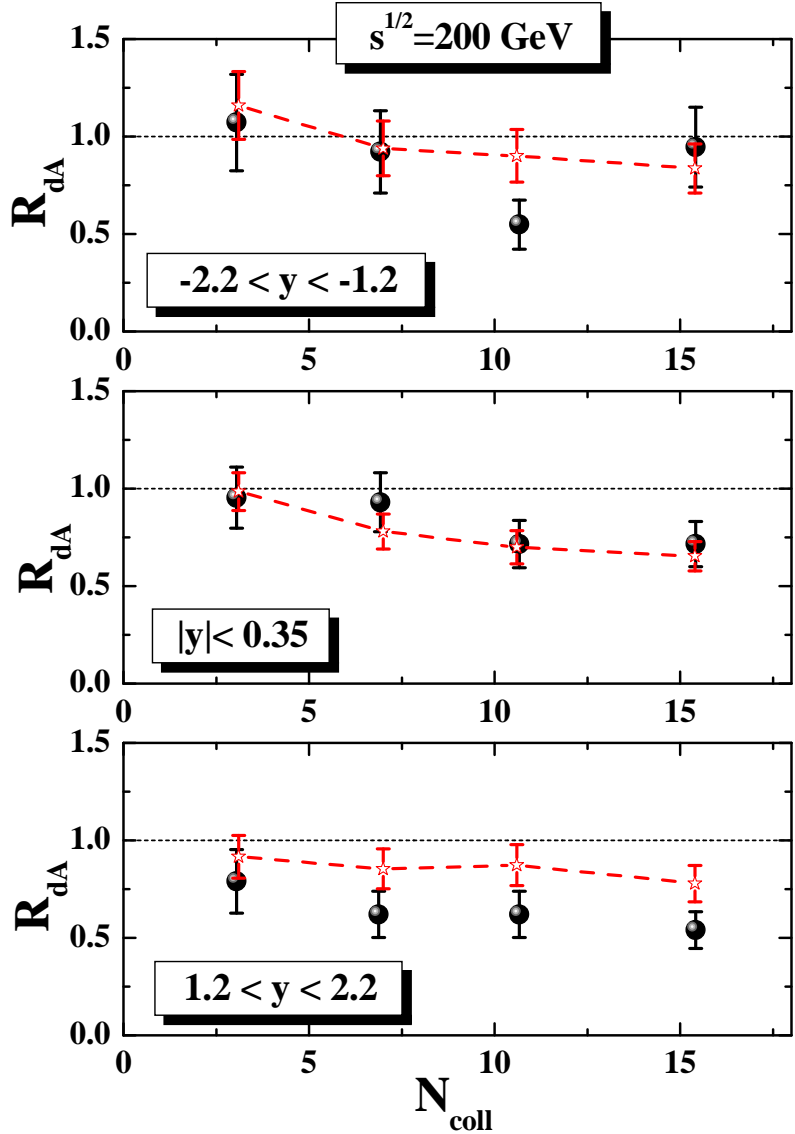,width=0.75\textwidth}}
\caption{The ratio $R_{dA}$ (\ref{FigRdA}) for backward, central
and forward rapidity bins as a function of the number of binary
collisions $N_{coll}$ for $d+Au$ at $\sqrt{s}$ = 200 GeV. The
experimental data have been taken from Ref. \cite{PHENIX08}. The
HSD results (stars connected by red dashed lines) show
calculations without including low-$x$ gluon shadowing and
slightly overestimate $R_{dA}$ in the forward interval 1.2 $< y <$
2.2. The theoretical error bars are due to the finite statistics
of the calculation.} \label{dAu2}
\end{figure}

It is seen from Fig.~\ref{dAu} that the calculations follow
approximately the decrease in $R_{dA}$ with rapidity, however,
with a tendency to overshoot at forward rapidity. Within error
bars we find  the values of  $\sigma_{c\bar c B}$ from
(\ref{sigmacB}) to be compatible with the inclusive RHIC
measurement as well as with the lower energy data~\cite{Borges}.
This finding is also in line with the analysis of the PHENIX
Collaboration in Ref. \cite{PHENIX08}

In order to shed some further light on the role of shadowing, we
compare our calculations for $R_{dA}$ in different rapidity bins
as a function of the centrality of the $d+Au$ collision, which in
Fig. \ref{dAu2} is represented by the number of binary collisions
$N_{coll}$. The latter number is directly taken from the number of
binary hard $NN$ collisions in the transport calculation while the
comparison with experiment is based on a Glauber model analysis of
the data similar to that performed in Ref. \cite{Gran}. The actual
results displayed in Fig. \ref{dAu2} (stars connected by dashed
lines) and the PHENIX data from Ref. \cite{PHENIX08} are roughly
compatible for the rapidity intervals -2.2 $< y <$ -1.2 and $|y|
<$ 0.35, but demonstrate that the suppression at forward rapidity
(1.2 $< y < $2.2) is underestimated in the color-dipole
dissociation model with a constant cross section of 4.18 mb. This
clearly points to the presence of shadowing effects at least at
forward rapidities which is not so pronounced in the inclusive
data set in Fig. \ref{dAu}. A more serious question is a
quantification of the shadowing due to the limited statistics of
both the experimental data and the calculations. Here we do not
attempt to attribute a fixed number for the shadowing effect but
merely point out that independent high statistics data will be
necessary to fix this unsatisfactory situation from the
experimental side.

Nevertheless, some note of caution is appropriate for the further
analysis of charmonium suppression in $Au+Au$ collisions: There
are `cold nuclear matter effects' such as `gluon shadowing' beyond
those incorporated in the transport calculations, and especially
quantitative statements about any `agreement with data' might have
to be reconsidered. In case of $Au+Au$ reactions the shadowing
from projectile/target will show up symmetrically around $y=0$ and
in part contribute to the stronger $J/\Psi$ suppression at
forward/backward rapidities. Nevertheless, following Granier de
Cassagnac \cite{Gran}, an anomalous suppression of $J/\Psi$ beyond
`cold nuclear matter effects' is clearly present in the $Au+Au$
data to be investigated below.

%********************************************************************
\section{Scenarios for the anomalous $J/\Psi$ suppression} \label{scenarios}

It is well known that the baryonic (normal) absorption alone
cannot explain the suppression of charmonia in heavy-ion
collisions with increasing centrality \cite{rev1}. We have implemented
in HSD several different mechanism
for the additional (anomalous) suppression of charmonia which will
be explained in the following Subsections. By comparing the
results from these scenarios to each other and to the available data
the mechanism of charmonium interactions with the medium can be
probed.

%*********************************
\subsection{`Comover' suppression (and recombination)} \label{comover}

First of all let us stress that the interactions with `comoving'
mesons lead not only to the dissociation of charmonia, but also
to their recreation via the inverse recombination process $D+\bar D
\to c \bar c + m$, where $m= \{ \pi, \rho, \omega, K, ... \}$.
 As already pointed out before, the $J/\Psi,
\chi_c, \Psi^\prime$ formation cross sections by open charm mesons
or the inverse `comover' dissociation cross sections are not well
known and the significance of these channels is discussed
controversely in the
literature~\cite{BMS,Rafelski,Bernd,Redlich,I2,I3,KoO}. We here
follow the concept of Refs.~\cite{brat03,brat04} and introduce a
simple 2-body transition model with a single parameter $|M_0|^2$,
that allows to implement the backward reactions uniquely by
employing detailed balance for each individual channel.

Since the charmonium-meson dissociation and backward reactions
typically occur with low relative momenta (`comovers'), it is
legitimate to write the cross section for the process $1+2\to 3+4$
as
\begin{equation}
\label{model}
 \sigma_{1+2\to 3+4}(s) = 2^4 \frac{E_1 E_2 E_3 E_4}{s}
|\tilde M_i|^2 \left(\frac{m_3+m_4}{\sqrt{s}}\right)^6
\frac{p_f}{p_i},
\end{equation}
 where $E_k$ denotes the energy of hadron $k$
$(k=1,2,3,4)$, respectively. The initial and final momenta for
fixed invariant energy  $\sqrt{s}$ are given by
\begin{eqnarray}
p_i^2 = \frac{(s-(m_1+m_2)^2)(s-(m_1-m_2)^2)}{4s}, \nonumber\\
p_f^2 = \frac{(s-(m_3+m_4)^2)(s-(m_3-m_4)^2)}{4s}, \label{moment}
\end{eqnarray}
where $m_k$ denotes the mass of hadron $k$. In (\ref{model})
$|\tilde M_i|^2$ ($i=\chi_c, J/\Psi, \Psi^\prime$) stands for the
effective matrix element squared, which for the different 2-body
channels is taken of the form
\begin{eqnarray}
&&\hspace*{-3mm}|\tilde M_i|^2 =|M_i|^2  \ \ {\rm for} \
    \ (\pi,\rho)+(c\bar c)_i \to D+\bar{D} \label{mod}\\
&&\hspace*{-3mm}|\tilde M_i|^2 = 3 |M_i|^2  \ \ {\rm for} \
    \ (\pi,\rho)+(c\bar c)_i  \to D^*+\bar{D}, \ D+\bar{D}^*, \ D^* + \bar{D}^* \nonumber\\
&&\hspace*{-3mm}|\tilde M_i|^2 = \frac{1}{3} |M_i|^2 \ \ {\rm
for}\
    \ (K,K^*)+(c\bar c)_i  \to D_s + \bar{D}, \ \bar{D}_s + D \nonumber \\
&&\hspace*{-3mm}|\tilde M_i|^2 =  |M_i|^2  \ \ {\rm for} \
    \ (K,K^*)+(c\bar c)_i  \to D_s + \bar{D}^*, \ \bar{D}_s + D^*,\ D^*_s + \bar{D}, \nonumber \\
&&\phantom{|\tilde M_i|^2 =  |M_i|^2  \ \ {\rm for} \ \
(K,K^*)+(c\bar c)_i  \to D_s + \bar{D}^*, \ }     \bar{D}^*_s + D,
\ \bar{D}^*_s + D^* \nonumber
\end{eqnarray}
The relative factors of 3 in (\ref{mod}) are guided by the sum
rule studies in~\cite{korean} which suggest that the cross section
is increased whenever a vector meson $D^*$ or $\bar{D}^*$ appears
in the final channel while another factor of 1/3 is introduced for
each $s$ or $\bar{s}$ quark involved. The factor $\left(
{(m_3+m_4)}/{\sqrt{s}} \right)^6 $ in (\ref{model}) accounts for
the suppression of binary channels with increasing $\sqrt{s}$ and
has been fitted to the experimental data for the reactions $\pi +
N \rightarrow \rho+N, \omega+N, \phi+N, K^+ +\Lambda$ in Ref.
\cite{CaKo}.

We use  the same matrix elements for the
dissociation of all charmonium states $i$ ($i=\chi_c, J/\Psi,
\Psi^\prime$) with mesons:
\begin{eqnarray}
 |M_{J/\Psi}|^2 = |M_{\chi_c}|^2 = |M_{\Psi^\prime}|^2 = |M_0|^2.
\label{MatrElem}
\end{eqnarray}
We note for completeness that in Ref.~\cite{brat03} the parameter $|M_0|^2$ was
fixed by comparison to the $J/\Psi$ suppression data from the NA38
and NA50 Collaborations for S+U and Pb+Pb collisions at 200 and
158 AGeV, respectively. In a later study~\cite{Olena}, however,
this parameter has been readjusted in accordance with the updated
value of the cross section~(\ref{sigmacB}) of charmonium
dissociation on baryons (following the latest NA50 and NA60
analysis~\cite{NA60,NA50pA}). The best fit is obtained for
$|M_0|^2=0.18$~fm$^2$/GeV$^2$; this value will be employed in our
following studies, too.

The advantage of the model introduced in \cite{brat03,brat04} is
that detailed balance for the binary reactions can be employed
strictly for each individual channel, {\it i.e.}
\begin{eqnarray}
\!\!\sigma_{3+4 \rightarrow 1+2}(s) =
 \sigma_{1+2 \rightarrow 3+4}(s)
\frac{(2S_1+1)(2S_2+1)}{(2S_3+1)(2S_4+1)} \ \frac{p_i^2}{p_f^2}, \
\label{balance}
\end{eqnarray}
 and the role of the backward reactions
($(c\bar c)_i$+meson formation by $D+\bar{D}$ flavor exchange) can
be explored without introducing any additional parameter once
$|M_0|^2$ is fixed. In Eq.~(\ref{balance}) the quantities $S_j$
denote the spins of the particles, while ${p_i^2}$ and ${p_f^2}$
denote the cms momentum squared in the initial and final channels,
respectively. The uncertainty in the cross sections
(\ref{balance}) is of the same order of magnitude as that in
Lagrangian approaches using {\it e.g.} $SU(4)_{flavor}$ symmetry
\cite{Konew,Ko}, since the form factors at the vertices are
essentially unknown~\cite{korean}. It should be pointed out that
the `comover' dissociation channels for charmonia are described in
HSD with the proper individual thresholds for each channel in
contrast to the more schematic `comover' absorption
model in Refs.~\cite{Capellanew,Capella}.

The regeneration of charmonia by recombination of $D$ ($D^*$)
mesons in the hadronic phase was first studied by C.M.~Ko and
collaborators in~\cite{KoO}. The conclusion at that time was that
this process is unlikely  at RHIC
energies~\cite{Redlich,KoO,Redlich2}. On the other hand, it has
been shown within HSD~\cite{brat03} that the contribution of the
$D+\bar D$ annihilation to the produced $J/\Psi$ at RHIC is
considerable. Moreover, the equilibrium in the reaction $J/\Psi+m
\leftrightarrow D\bar D$ is reached (i.e. the charmonium recreation is
comparable with the dissociation by `comoving' mesons).
The reason for such differences is that the
pioneering study~\cite{KoO} within the hadron gas model was
confined to $J/\Psi$ reactions with $\pi$'s into two particular
$D\bar D$ channels ($D+\bar D^*$ and $D^*+\bar D^*$). On the
contrary, in Ref.~\cite{brat03} the interactions with all mesons into
all possible combinations of $D\bar D$ states have been taken into
account. Note that the $\rho$-meson density at RHIC is large such
that  the channel with the most abundant $\rho$-meson resonance is
dominant. Furthermore, in Ref.~\cite{brat03} the feed down from
$\chi _c$ and $\Psi'$ decays has been considered. The results
of~\cite{brat03} are in accordance with independent studies in
Refs.~\cite{DD1,DD2,DD3,DD4}. Later work within the HSD
approach~\cite{Olena2} has supported the conclusions
of Ref.~\cite{brat03} and stressed the importance for $D\bar D$
annihilation in the late (purely hadronic) stages of the collisions.

%*********************************
\subsection{`Threshold melting'}

This scenario is based on the idea of sequential
dissociation of charmonia with increasing temperature
\cite{Satz,Satznew,Satzrev,KSatz}, {\it i.e.} of
charmonium melting in the QGP due to color screening as soon as the
fireball temperature reaches the dissociation temperatures of
($\approx 2T_c$ for $J/\Psi$, $\approx T_c$ for excited states,
where $T_c$ stands for the critical temperature of the deconfinement
phase transition). In the early approaches the temperature of the fireball
has been estimated using e.g. the Bjorken formula (\ref{bjorken}). We modify the
standard sequential dissociation model in two aspects: (i) the
energy density is calculated locally and microscopically instead
of using schematic estimates ({\it cf.}
section~\ref{energydensity}); (ii) the model incorporates a
charmonium regeneration mechanism (by $D\bar D$ annihilation
processes).

The `threshold scenario' for charmonium dissociation now is
implemented in a straight forward way: whenever the local energy
density $\varepsilon(x)$ is above a threshold value
$\varepsilon_j$, where the index $j$ stands for $J/\Psi, \chi_c,
\Psi^\prime$, the charmonium is fully dissociated to $c +
\bar{c}$. The default threshold energy densities adopted are
\be
\mbox{ $\varepsilon_{J/\Psi} = 16$ GeV/fm$^3$ ,
$\varepsilon_{\chi_c} = 2$ GeV/fm$^3$, and
$\varepsilon_{\Psi^\prime} =2 $ GeV/fm$^3$.} \label{melt} \ee

The dissociation of charmonia is widely studied using lattice
QCD (lQCD) \cite{Aarts,Petreczky,lQCD1,lQCD2,lQCD3} in order to determine
the dissociation temperature (or energy density) via the maximum entropy method.
On the other hand one may use potential models - reproducing the charmonium
excitation spectrum in vacuum - to calculate Mott transition temperatures
in a hot medium. Both
approaches have their limitations and the quantitative agreement
between the different groups is still unsatisfactory:
\begin{itemize}
\item (A) Potential models employ the static heavy quark-antiquark
pair free energy - calculated on the lattice - to obtain the charmonium
spectral functions. This leads to the dissociation
temperatures~\cite{Mocsy} $$T_{melt}(J/\Psi) \le 1.2 \, T_c, \
T_{melt}(\chi_{c}) \le T_c, \ T_{melt}(\Psi') \le T_c .$$
\item (B) The maximum entropy method is used to relate the
Euclidean thermal correlators of charmonia - calculated on the
lattice - to the corresponding spectral functions and yields
higher dissociation temperatures~\cite{Aarts} $$T_{melt}(J/\Psi) =
1.7\! -\! 2 \, T_c, \ T_{melt}(\chi_{c}) = 1.1\! - \! 1.2 \, T_c$$
or~\cite{Petreczky} $$T_{melt}(J/\Psi) \ge 1.5 \, T_c, \
T_{melt}(\chi_{c})=1.1 \, T_c.$$
\end{itemize}
Our earlier analysis of  experimental data at the SPS in the `threshold
melting' approach~\cite{Olena} lead us to conclude from the
observation of a considerable amount of $J/\Psi$ in the most
central $Pb+Pb$ collisions  that the assumption of a melting of $J/\Psi$ close
to $T_c$ contradicts the data. Therefore, the values (\ref{melt})
are applied also in the current study.

%*********************************
\subsection{Discriminating hadronic and partonic phases}

Two more scenarios are implemented in our present HSD simulations
that are closely related to the `comover suppression' and the
`threshold melting' scenarios outlined in the previous
sub-sections. The essential difference is that the comoving
hadrons (including the $D$-mesons) exist only at energy densities
below some  energy density $\varepsilon _{cut}$, which is a free
parameter. We employ $\varepsilon _{cut}=\varepsilon_c \approx
1$~GeV/fm$^3$, which is equal to the critical energy density
$\varepsilon_c$ for the parton/hadron phase transition. This
scenario clearly separates `formed hadrons' from possible
pre-hadronic states at higher energy densities. Indeed, it is
currently not clear whether  $D$- or $D^*$-mesons survive at
energy densities above $\varepsilon_c$ but hadronic correlators
with the quantum numbers of the hadronic states are likely to
persist above the phase transition~\cite{Rapp05}. One may
speculate that similar correlations (pre-hadrons) survive also in
the light quark sector above $T_c$ such that `hadronic comovers'
-- with modified spectral functions -- might show up also at
energy densities above $\varepsilon_c$.

We recall that the concept of (color neutral) pre-hadrons -
explained in more detail in Refs. \cite{Falter1,HPT1} - has been
also used in the hadron electroproduction studies off nuclei in
Refs. \cite{Falter1,Falter2} as well as for high $p_T$ hadron
suppression \cite{HPT1} or jet suppression at RHIC energies
\cite{HPT2}. It has been found that the pre-hadron concept works
well for hadron attenuation in nuclei at HERMES energies
\cite{Falter1,Falter2} but underestimates the high $p_T$ hadron
suppression \cite{HPT1} as well as the jet attenuation at RHIC
energies \cite{HPT2}. Nevertheless, the amount of attenuation due
to such pre-hadronic interactions emerged to be about 50\% of the
experimentally observed suppression at RHIC such that their effect
might not simply be discarded. It should be stressed that the
concept of pre-hadrons refers to the string breaking mechanism as
described in Refs. \cite{Falter1,HPT1} and is independent on the
energy density. A detailed study on the space-time evolution of
pre-hadrons and their formation to hadrons for $pp$ collisions has
been performed by Gallmeister and Falter in Ref. \cite{Fal}.

In line with the investigations in Refs. \cite{HPT1,HPT2} we also
study $J/\Psi$ production and absorption in $Au+Au$ collisions at
$\sqrt{s}=200$~AGeV assuming  the absorption of charmonia on
pre-hadrons as well as their regeneration by pre-hadrons. This
adds additional interactions of the particles with charm quarks
(antiquarks) in the very early phase of the nucleus-nucleus
collisions as compared to the default HSD approach. Since these
pre-hadronic (color-dipole) states represent some new
degrees-of-freedom, the interactions of charmed states with these
objects have to be specified separately.

For notation we define a pre-hadronic state consisting of a
quark-antiquark pair as pre-meson ${\tilde m}$ and a state consisting
of a diquark-quark pair as pre-baryon ${\tilde B}$. The dissociation
cross section of a $c{\bar c}$ color dipole state with a pre-baryon is
taken to be of the same order as with a formed baryon,
\begin{equation} \label{sss1}
\sigma_{c{\bar c}{\tilde B}}^{diss} = 5.8 \ {\rm mb},
\end{equation}
whereas the cross section with a pre-meson follows
from the additive quark model as \cite{Falter1,Falter2}
\begin{equation} \label{sss2}
\sigma_{c{\bar c}{\tilde m}}^{diss} = \frac{2}{3} \sigma_{c{\bar c}{\tilde B}}^{diss}.
\end{equation}
Elastic cross sections are taken as
\begin{equation} \label{sss3}
\sigma_{c{\bar c}{\tilde B}}^{el} = 1.9 \ {\rm mb}, \hspace{2cm}
 \sigma_{c{\bar c}{\tilde m}}^{el}
= \frac{2}{3} \sigma_{c{\bar c}{\tilde B}}^{el}.
\end{equation}
Furthermore, elastic interactions of a charm quark (antiquark) are
modeled by the scattering of an unformed $D$ or $D^*$ meson on
pre-hadrons with only light quarks as
\begin{equation} \label{sss4}
\sigma_{D{\tilde B}}^{el} = 3.9 \ {\rm mb}, \hspace{2cm}
 \sigma_{D {\tilde m}}^{el}
= \frac{2}{3} \sigma_{D{\tilde B}}^{el}.
\end{equation}
In this way we may incorporate in HSD some dynamics of
quark-antiquark pairs with a medium that has not yet formed the
ordinary hadrons. However, it has to be stressed
 that further explicit partonic degrees of freedom, i.e.
gluons and their mutual interactions as well as gluon interactions
with quarks and antiquarks, are not taken into account in the
present HSD approach. Therefore, we do not expect to reproduce any
details of the measured $J/\Psi$ yield. The study of this
particular model situation is motivated first of all by the
possibility to assess the conceptual influence of charm
scattering on pre-hadrons (in the early reaction phase)
on the final rapidity distribution of
the $J/\Psi$'s (see below).

%*********************************************************************
\section{Comparison to data} \label{results}

\begin{figure}
%\phantom{a}
%\vspace*{-0.5cm}
\centerline{\psfig{figure=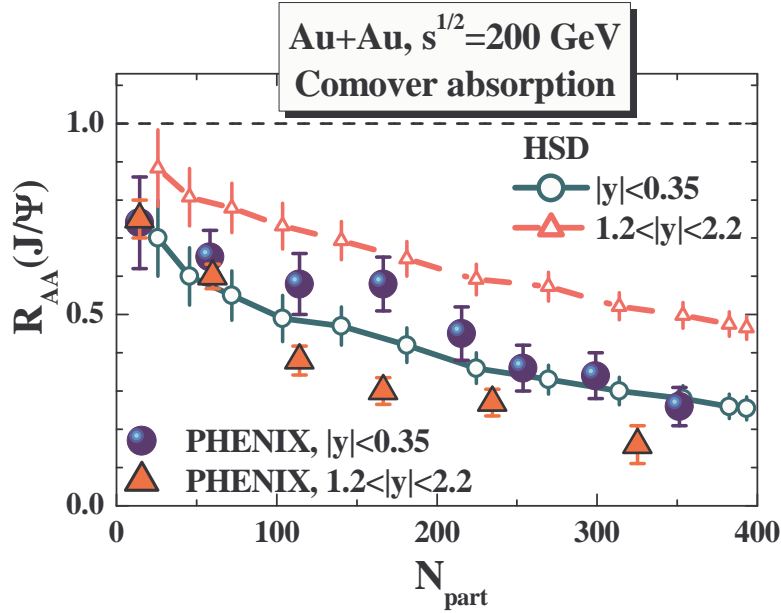,width=0.75\textwidth}}
\caption{The $J/\Psi$ nuclear modification factor $R_{AA}$
(\ref{raa}) for Au+Au collisions at $\sqrt{s} = 200$~AGeV as a
function of the number of participants $N_{part}$ in comparison to
the data from~\cite{PHENIXNov06} for midrapidity (full circles)
and forward rapidity (full triangles). The HSD results for the
purely hadronic `comover' scenario are displayed  in terms of the
lower (green solid) line with open circles for midrapidity
$J/\Psi's$ ($|y| \leq 0.35$) and in terms of the upper (red
dashed) line with open triangles for forward rapidity ($1.2 \leq |y|
\leq 2.2$). } \label{xs5}
\end{figure}

In the transport approach we calculate the $J/\Psi$ survival
probability $S_{J/\Psi}$ and the nuclear modification factor
$R_{AA}$ as
\begin{equation} S_{J/\Psi}  =
\frac{N^{J/\Psi}_{fin}}{N^{J/\Psi}_{BB}}, \mbox{ } \ \
R_{AA} = \frac{d N^{J/\Psi}_{AA} / d y }{N_{coll} \cdot
d N^{J/\Psi}_{pp} /d y},
\label{raa}
\end{equation}
where $ N^{J/\Psi}_{fin}$ and $N^{J/\Psi}_{BB}$ denote the final
number of $J/\Psi$ mesons and the number of $J/\Psi$'s produced
initially by $BB$ reactions, respectively. Note that $
N^{J/\Psi}_{fin}$ includes the decays from the final $\chi_c$. In
(\ref{raa}),  $d N^{J/\Psi}_{AA} / d y$ denotes the final yield of
$J/\Psi$ in $A A$ collisions, $d N^{J/\Psi}_{pp} / d y$  is the
yield in elementary $p p$ reactions while  $N_{coll}$ is the
number of initial binary collisions.

\begin{figure}
%\phantom{a}
%\vspace*{-0.5cm}
\centerline{\psfig{figure=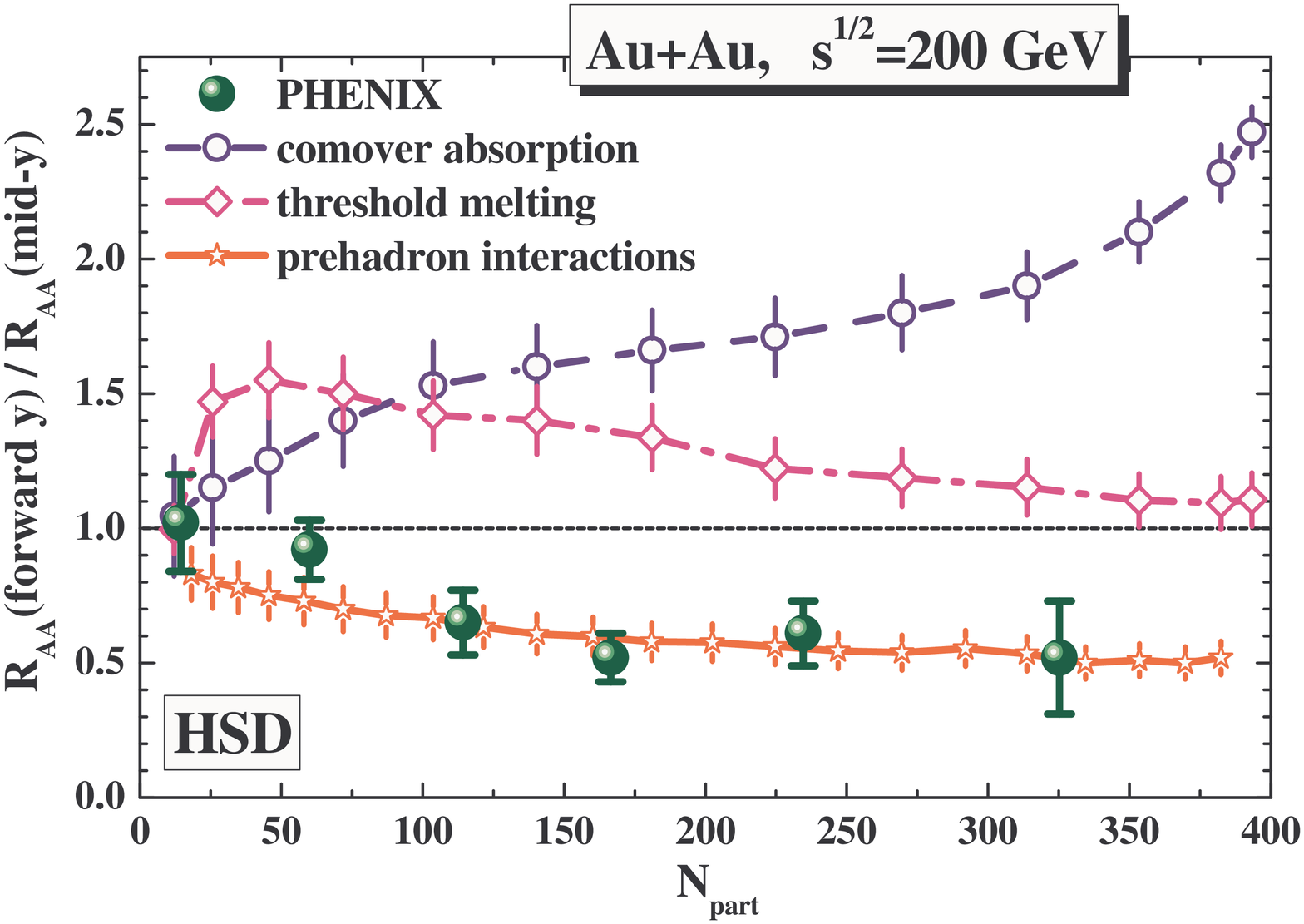,width=0.75\textwidth}} \caption{The
ratio of the nuclear modification factors $R_{AA}$ at mid-rapidity
($|y|<0.35$) and at forward rapidity ($1.2<|y|<2.2$) {\it vs}
centrality in $Au+Au$ collisions at $\sqrt{s}=200$~GeV. The HSD results
in the purely hadronic scenario (`comover absorption') are displayed in
terms of the blue dashed line (with open circles) and in case of the
`threshold melting' scenario in terms of the violet dot-dashed  line
(with open squares).  The error bars on the theoretical results
 indicate the statistical uncertainty due to the finite number of
Monte-Carlo events in the calculations. The lower full green dots
represent the data of the PHENIX Collaboration~\cite{PHENIXNov06}.
Note that the data have an additional systematic uncertainty of $\pm 14
\%$. The lower solid (red) line with stars gives the result for the
'comover absorption' scenario when including additional pre-hadronic
interactions with charm (see text).}
\label{Ratio}
\end{figure}

The suppression of charmonia by the `comover' dissociation channels
within the model described in~\cite{Olena} for a matrix element squared
$|M_0|^2$ = 0.18 fm$^2$/GeV$^2$ has been presented already in
Ref.~\cite{Olena2} as well as the results for the `threshold melting  scenario'
employing the thresholds $\varepsilon_{J/\Psi} = 16$ GeV/fm$^3$,
$\varepsilon_{\chi_c} = \varepsilon_{\Psi^\prime} =2$ GeV/fm$^3$. Note
that the charmonium reformation channels by $D+\bar{D}$ channels had
been incorporated, too ({\it cf.} Ref.~\cite{brat03}). Since the PHENIX
Collaboration has released a new data set we compare our calculations
with the most recent PHENIX data~\cite{PHENIXNov06} in Fig. \ref{xs5}
for the $J/\Psi$ nuclear modification factor $R_{AA}$ (\ref{raa}) for
Au+Au collisions at $\sqrt{s} = 200$~AGeV as a function of the number
of participants $N_{part}$ for midrapidity (full circles) and forward
rapidity (full triangles). The HSD results for the purely hadronic
`comover' scenario are displayed in terms of the lower (blue solid)
 line with open circles for midrapidity $J/\Psi's$ ($|y| \leq 0.35$)
and in terms of the upper (red dashed) line with open triangles for
forward rapidity ($1.2 \leq |y| \leq 2.2$). The numerical results appear
acceptable at midrapidity ($|y| \leq 0.35$) but the even larger
suppression at forward rapidity (seen experimentally) is fully missed
(cf. Ref.~\cite{Olena2}).

The failure of the traditional 'comover absorption' model as well as
'threshold melting' scenario at the top RHIC energy is most clearly
seen in the centrality dependence of the ratio of the nuclear
modification factors $R_{AA}$ at forward rapidity ($1.2<|y|<2.2$) and
at mid-rapidity ($|y|<0.35$) as shown in Fig.  \ref{Ratio}. The HSD
results in the purely hadronic scenario (`comover absorption') are
displayed in terms of the blue dashed line (with open circles) and in
case of the `threshold melting' scenario in terms of the dot-dashed
violet line (with open squares).  The error bars on the theoretical
results indicate the statistical uncertainty due to the finite number
of Monte-Carlo events in the calculations. The lower full green dots in
Fig. \ref{Ratio} represent the corresponding data of the PHENIX
Collaboration~\cite{PHENIXNov06} which show a fully different pattern
as a function of centrality (here given in terms of the number of
participants $N_{part}$). The failure of these 'standard' suppression
models at RHIC has lead to the conclusion in Ref. \cite{Olena2} that
the hadronic 'comover absorption and recombination' model is falsified
by the PHENIX data and that strong interactions in the pre-hadronic (or
partonic) phase should be necessary to explain the large suppression at
forward rapidities.

\begin{figure}
%\phantom{a}
%\vspace*{-0.5cm}
\centerline{\psfig{figure=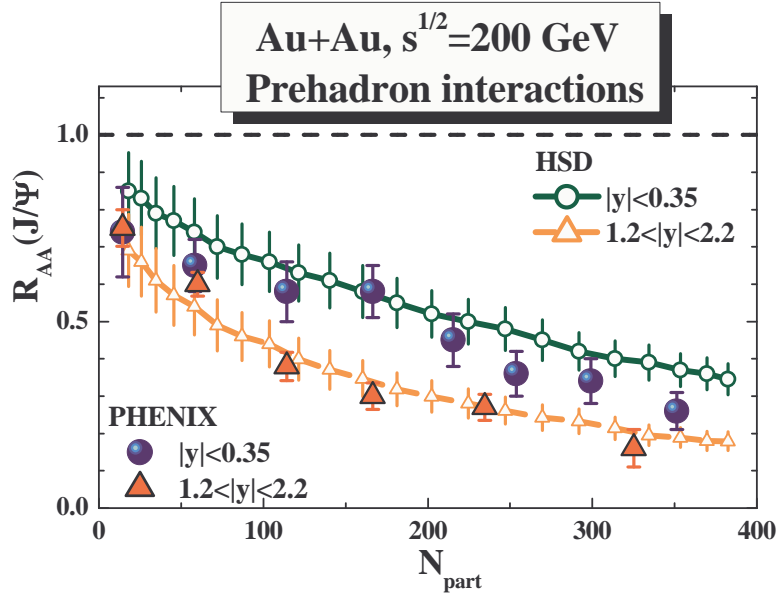,width=0.75\textwidth}}
\caption{The $J/\Psi$ nuclear modification factor $R_{AA}$
(\ref{raa}) for Au+Au collisions at $\sqrt{s} = 200$~AGeV as a
function of the number of participants $N_{part}$ in comparison to
the data from~\cite{PHENIXNov06} for midrapidity (full circles)
and forward rapidity (full triangles). The HSD results for the
hadronic `comover' scenario including additionally pre-hadronic
interactions of charm according to (\ref{sss1}) - (\ref{sss4})
 are displayed in terms of the upper (green solid) line
with open circles for midrapidity $J/\Psi's$ ($|y| \leq 0.35$) and
in terms of the lower (orange dashed) line with open triangles for
forward rapidity ($1.2 \leq |y| \leq 2.2$). } \label{xs5b}
\end{figure}

In this work we follow up the latter idea and incorporate in the
'comover scenario' the additional pre-hadronic cross sections
(\ref{sss1}) - (\ref{sss4}) for the early charm interactions to have a
first glance at the dominant effects. The $J/\Psi$ suppression pattern
in this case is shown in Fig. \ref{xs5b} in comparison to the same data
as in Fig.  \ref{xs5}. Now, indeed, the suppression pattern for central
and forward rapidities becomes rather similar to the data within the
statistical accuracy of the calculations. Indeed, the ratio of $R_{AA}$
at forward rapidity to midrapidity now follows closely the experimental
trend as seen in Fig. \ref{Ratio} by the lower red solid line.

\begin{figure}
%\phantom{a}
%\vspace*{-0.5cm}
\centerline{\psfig{figure=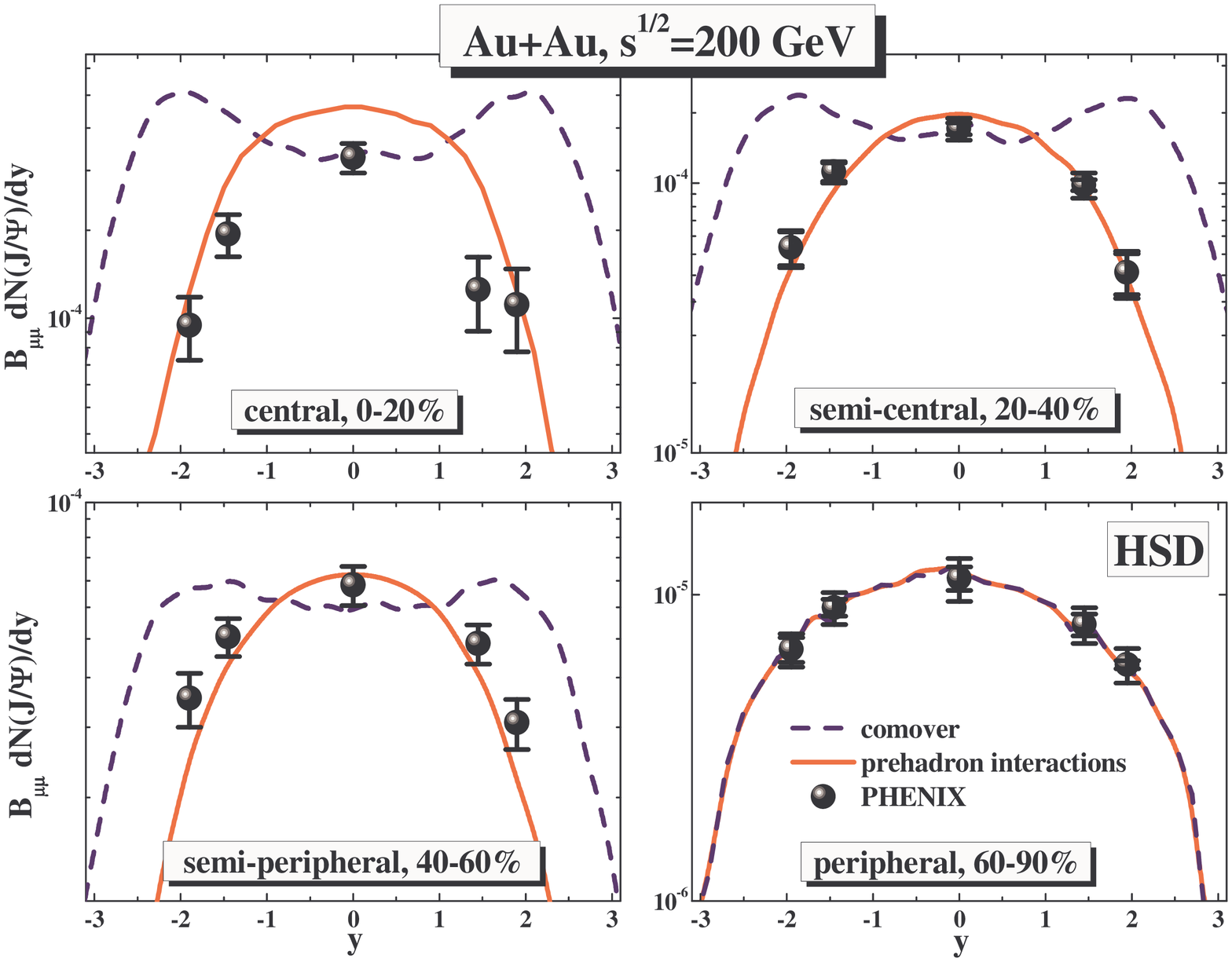,width=\textwidth}}
%\centerline{\psfig{figure=y1.eps,width=6cm},\psfig{figure=y2.eps,width=6cm}}
\caption{The rapidity distribution $dN_{J/\Psi}/dy$ for
different centralities  from the standard `comover' model
(dashed blue lines) and the 'comover' model with additional pre-hadronic
interactions of charm  according to (\ref{sss1}) - (\ref{sss4}) (solid red lines).
The full dots show the respective data from the PHENIX
Collaboration~\cite{PHENIXNov06}. The calculated lines have been
smoothed by a spline algorithm.  The reactions are Au+Au at
$\sqrt{s}$ = 200 GeV.} \label{y1}
\end{figure}

Some further information may be gained from the $J/\Psi$ rapidity
distributions in  Au+Au collisions at RHIC. The latter distribution is
shown in Fig. \ref{y1} in comparison to the PHENIX data for central
collisions (upper l.h.s.), semi-central (upper r.h.s.), semi-peripheral
(lower l.h.s.)  and peripheral reactions (lower r.h.s.) for the
standard 'comover' scenario (dashed blue lines) and the 'comover' model
including additionally pre-hadronic interactions of charm  according to
(\ref{sss1}) - (\ref{sss4}) (solid red lines).  Whereas for peripheral
reactions these additional early interactions practically play no role,
the latter lead to a narrowing of the $J/\Psi$ rapidity distribution
with the centrality of the collision (roughly in line with the data).
In the standard 'comover' model an opposite trend is seen: here the
interactions of charmonia with formed hadrons produce a dip in the
rapidity distribution at $y \approx$ 0 which increases with centrality
since the density of formed hadrons increases accordingly around
midrapidity. Since the total number of produced $c{\bar c}$ pairs is
the same (for the respective centrality class) and detailed balance is
incorporated in the reaction rates we find an surplus of $J/\Psi$ at
more forward rapidities. The net result is a broadening of the $J/\Psi$
rapidity distribution with centrality opposite to the trend observed in
experiment.

Summarizing the results displayed in Figs. 7 - 10 we like to point
out that the hadronic 'comover' dynamics for charmonium
dissociation and recreation - as well as the standard charmonium
'melting' scenario - do not match the general dependences of the
$J/\Psi$ in rapidity and centrality as seen by the PHENIX
Collaboration.  In fact, a  narrowing of the $J/\Psi$ rapidity
distribution cannot be achieved by comover interactions with
formed hadrons since the latter appear too late in the collision
dynamics. Only when including early pre-hadronic interactions with
charm a dynamical narrowing of the charmonium rapidity
distribution with centrality can be achieved as demonstrated more
schematically within our pre-hadronic interaction model.
Consequently, the PHENIX data on $J/\Psi$ suppression demonstrate
the presence and important impact of pre-hadronic or partonic
interactions in the early charm dynamics. This finding is line
with earlier studies in Refs. \cite{brat03,HPT1,HPT2}
demonstrating the necessity of non-hadronic degrees of freedom in
the early reaction phase for the elliptic flow $v_2$, the
suppression of hadrons at high transverse momentum $p_T$ and
far-side jet suppression in central Au+Au collisions at RHIC
energies.

%*************************************************************************
\section{Predictions, excitation functions and comparison to the statistical
hadronization model} \label{excitation}

In this Section we continue with predictions for future
measurements as well as model comparisons in order to allow for an
experimental discrimination between the model concepts.

\subsection{$\Psi '$ as an independent probe}

\begin{figure}
%\phantom{a}
%\vspace*{-0.5cm}
\centerline{\psfig{figure=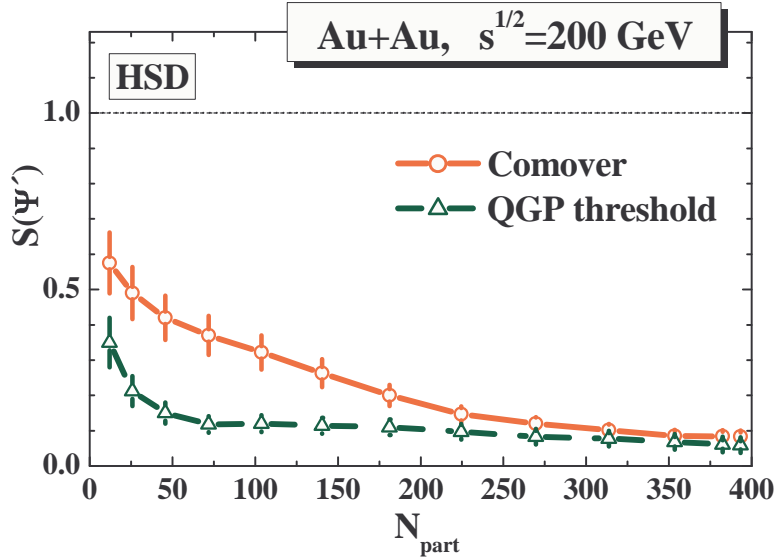,width=0.75\textwidth}}
\caption{Survival probability of $\Psi '$ in $Au+Au$ reactions at
$\sqrt{s}=200$~GeV in the `threshold melting' (dashed line with
triangles) and `comover' suppression (solid line with circles)
approaches, see text for details.} \label{SPP}
\end{figure}

As pointed in Ref. \cite{Olena} an independent measurement of
$\Psi^\prime$ will provide further information on the charm
reaction dynamics and final charmonium formation. For instance, a
leveling off of the $\Psi '$ to $J/\Psi$
ratio with increasing centrality would be a signal for
charm chemical equilibration in the medium \cite{PBM07,BMS,BMS2}.
Additionally, it provides a very clear distinction between the
`threshold melting' scenario and the `comover' approach. Since detailed
predictions for the $\Psi^\prime$ to $J/\Psi$ ratio as a function
of centrality have already been presented in Ref. \cite{Olena}
for FAIR and SPS energies we here complement the latter studies
by results for the  top RHIC energy although the suppression of $\Psi '$
mesons has not yet been measured at
RHIC.

In Fig~\ref{SPP}, we accordingly present the $\Psi '$ survival probability
$S_{\Psi '}$ defined as
\begin{equation} \label{sup} S_{\Psi '} =
\frac{N^{\Psi '}_{fin}}{N^{\Psi '}_{BB}},
\end{equation}
for $Au+Au$ at $\sqrt{s}=200$~GeV. In equation (\ref{sup}), $ N^{\Psi
'}_{fin}$ and $N^{\Psi '}_{BB}$ denote the number of final $\Psi '$
mesons and of those produced initially by $BB$ reactions, respectively.
One can see from Fig~\ref{SPP} that the `threshold melting' scenario at
RHIC predicts an almost complete melting of $\Psi '$, while a hadronic
`comover' absorption scenario shows a gradual decrease of the number of
$\Psi '$ with $N_{part}$. Similar differences between the models have
also been found at SPS energies \cite{Olena} where the presently
available data sets clearly favor the 'comover' model. On the other
hand our predictions for the top RHIC energy imply that the
$\Psi^\prime$ signal will be very low for mid-central and central Au+Au
collisions such that actual measurements will turn out to be very
demanding.

\subsection{Testing the assumption of statistical hadronization}
\label{statistical}

\begin{figure}
\begin{minipage}[l]{0.505\textwidth}
\psfig{figure=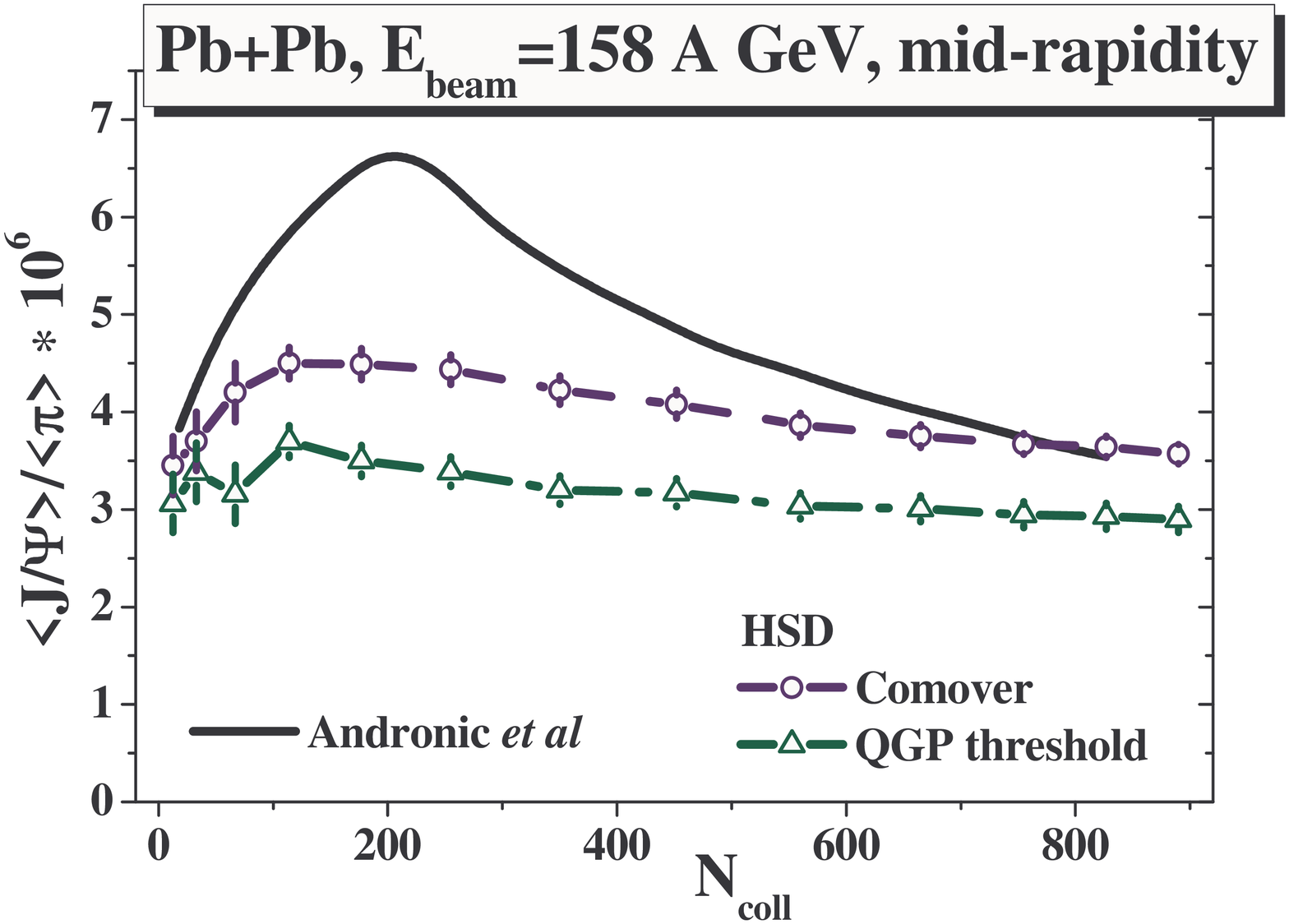,width=\textwidth}
\end{minipage}
\begin{minipage}[l]{0.495\textwidth}
\psfig{figure=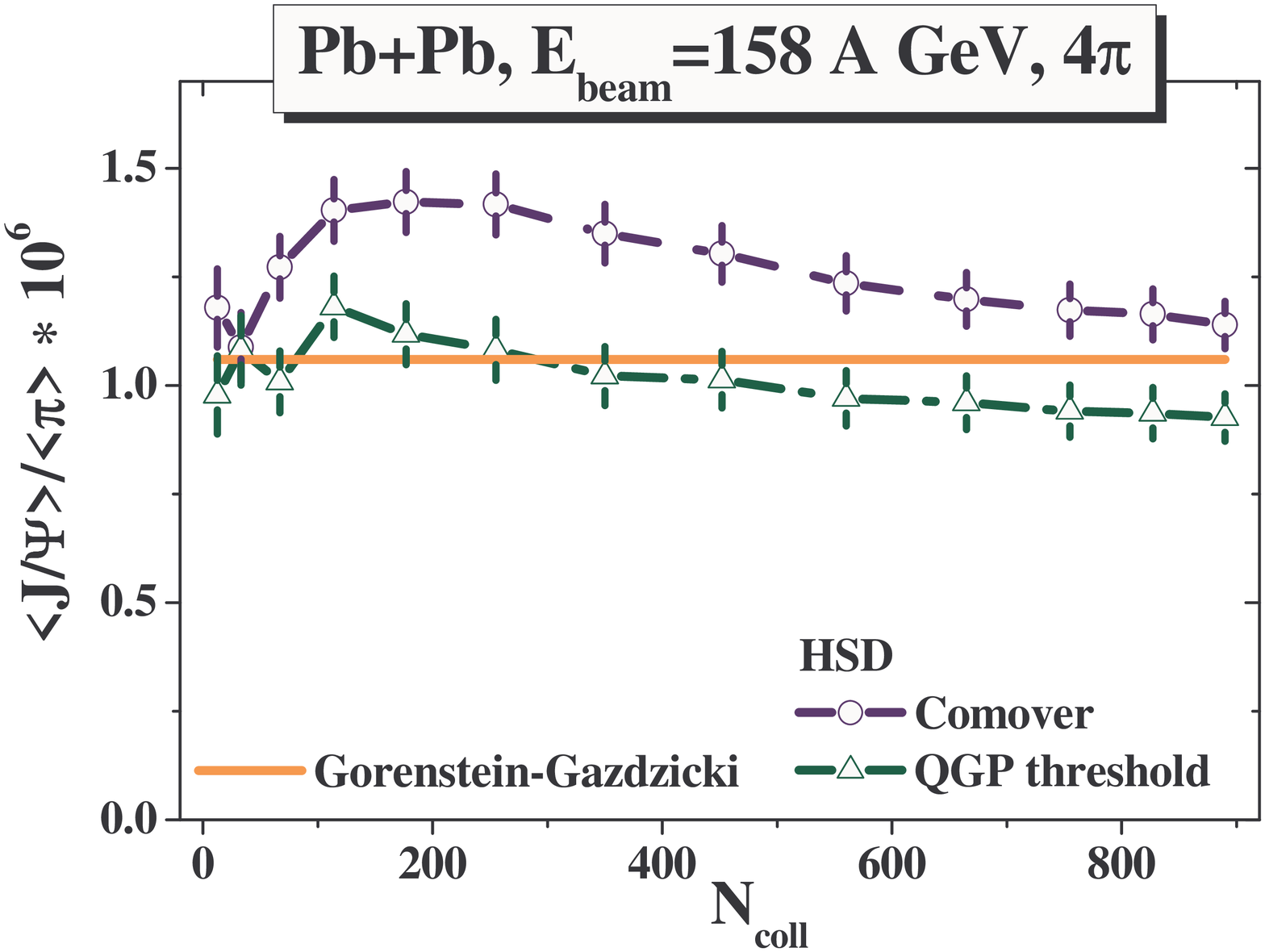,width=\textwidth}
\end{minipage}
\caption{Ratio of the averaged $J/\Psi$ to $\pi$ multiplicity  for
$Pb+Pb$ at the SPS beam energy of 158 A$\cdot$GeV at mid-rapidity
(l.h.s.) and in  full $4\pi$ acceptance (r.h.s.) as a function
of the number of binary collisions $N_{coll}$ for the different
suppression scenarios implemented in HSD - the `comover' model
(dashed blue line with open circles) and  the `threshold melting'
scenario (green dot-dashed line with open triangles) - in
comparison to the statistical model by Gorenstein and
Gazdzicki~\cite{MG**2} (r.h.s.; straight orange line) and the
statistical hadronization model by Andronic {\it et al.}
\cite{PBM07} (l.h.s.; solid black line). } \label{PsiPiSPS}
\end{figure}

\begin{figure}
%\phantom{a}
%\vspace*{-0.5cm}
\begin{minipage}[l]{0.5\textwidth}
\psfig{figure=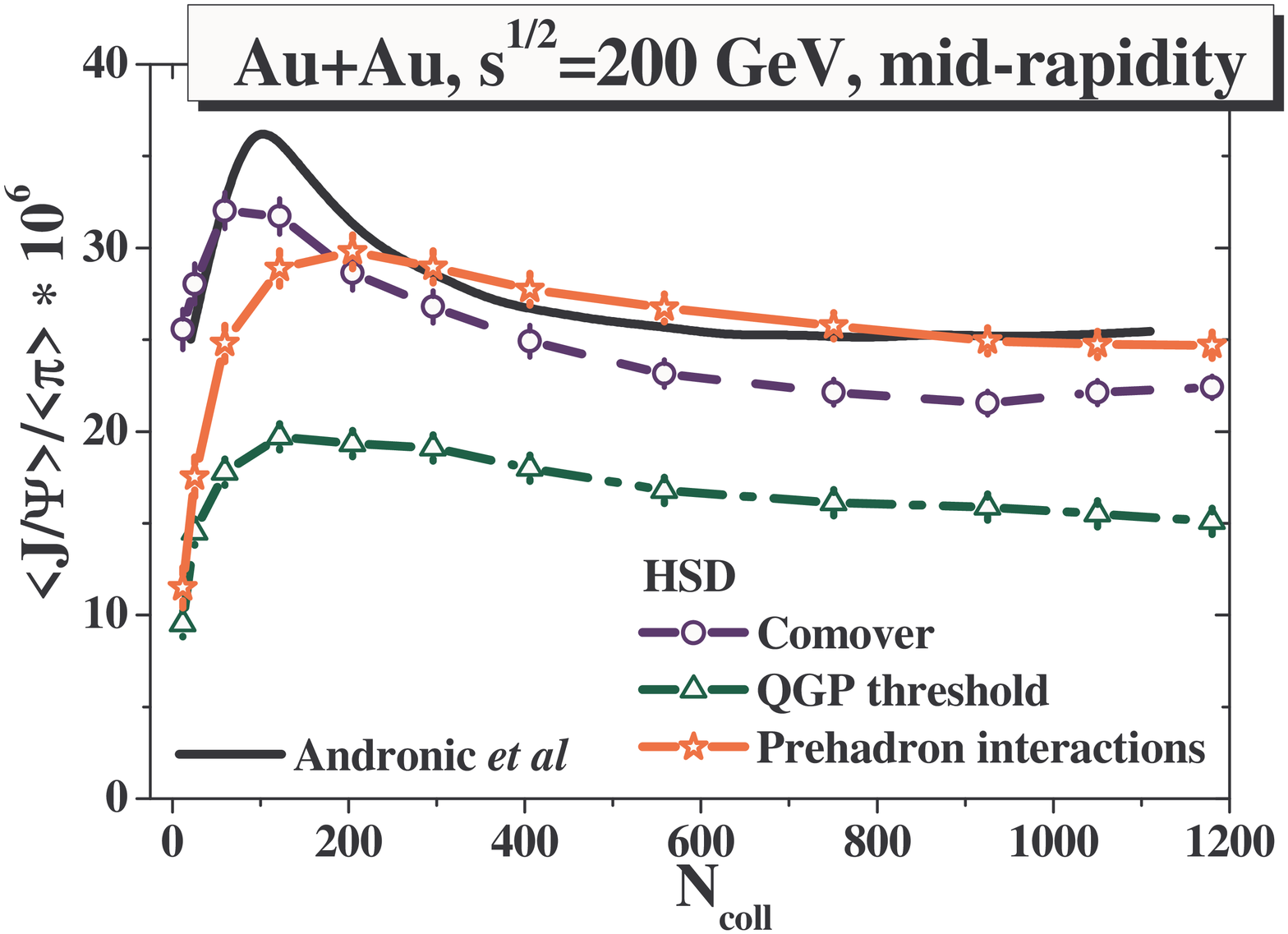,width=\textwidth}
\end{minipage}
\begin{minipage}[l]{0.5\textwidth}
\psfig{figure=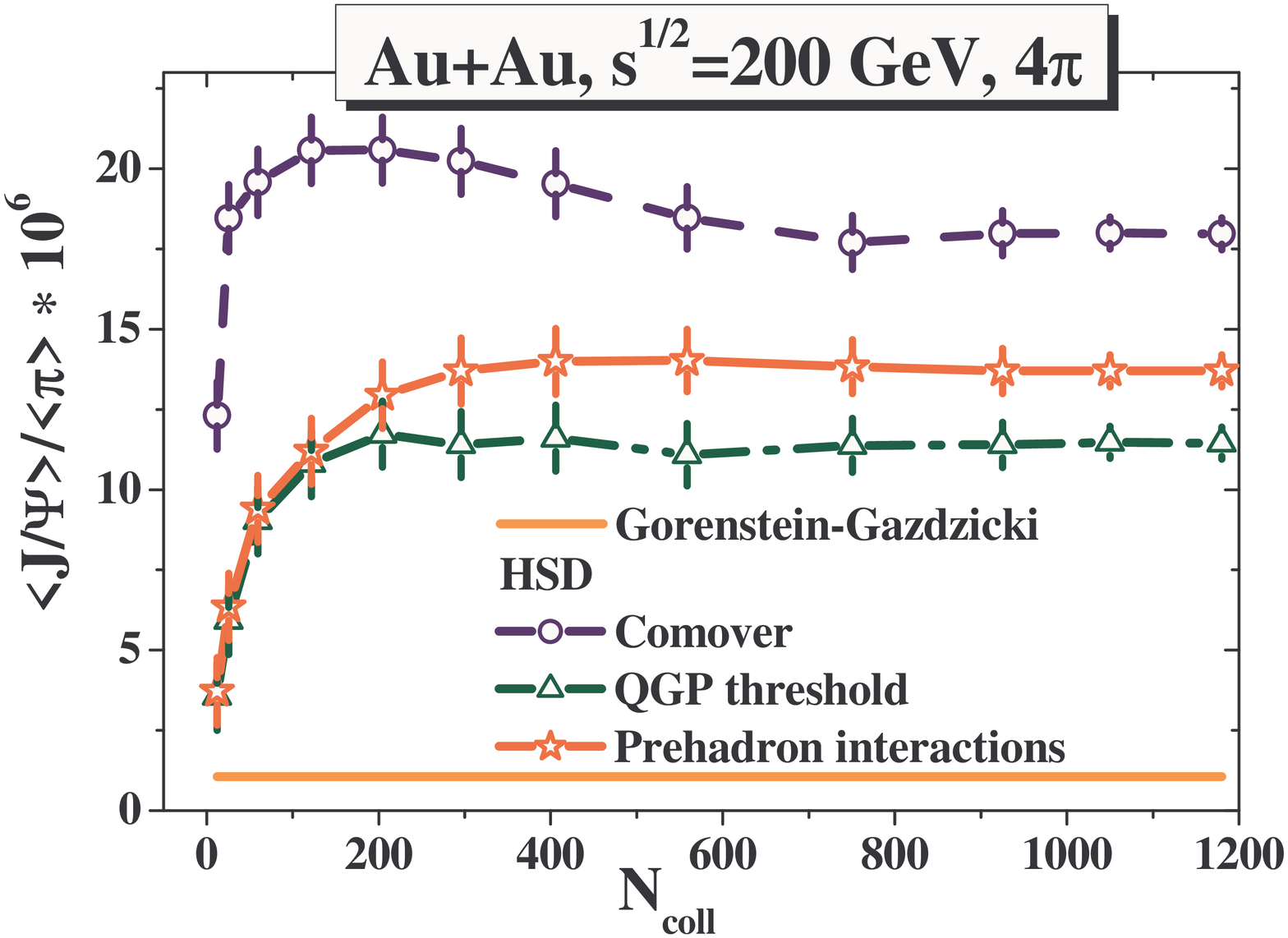,width=\textwidth}
\end{minipage}
\caption{Same as Fig.~\ref{PsiPiSPS} but for $Au+Au$ at the top RHIC
energy of $\sqrt{s}$ = 200 GeV. The red solid line
shows additionally the result of the 'comover' model including the
pre-hadronic charm interactions (see text).} \label{PsiPiRHI}
\end{figure}

The assumption of statistical hadronization -- {\it i.e.}  of
$J/\Psi$'s being dominantly produced at hadronization in a purely
statistically fashion according to available phase space and the
number of available $c$ and ${\bar c}$ quarks -- leads to a
scaling of the $\langle J/\Psi\rangle/\langle h\rangle$ ratio with
the system size~\cite{MG**2}, where $\langle h \rangle$ is the
average hadron multiplicity. Since $\langle h \rangle \sim
\langle\pi\rangle$, we calculate the ratio $\langle
J/\Psi\rangle/\langle\pi\rangle$ in HSD in the different scenarios
for charmonium suppression:
\begin{itemize}
\item `threshold melting' + recombination via $D\bar D\to c\bar c +m$
including the backward reactions $c \bar c + m \to D \bar D $, %
\item
hadronic (`comover') absorption: $D \bar D \to c \bar c  +
m$ and the backward reactions $c \bar c + m \to D \bar D $;
\item
`prehadron interactions': $D \bar D \to c \bar c  + m$ and the
backward reactions $c \bar c + m \to D \bar D $ as well as early
pre-hadronic charm interactions as described in Section 6.
\end{itemize}

The results of our calculations are shown in Fig.~\ref{PsiPiSPS}
together with the prediction of the statistical model of
Gorenstein and Gazdzicki ~\cite{MG**2} for the full phase space
(straight orange  line; r.h.s.) and the statistical hadronization
model by Andronic {\it et al.} \cite{PBM07,AndronicPrivat} for
mid-rapidity (solid black line; l.h.s.) for Pb+Pb at 158
A$\cdot$GeV. The centrality dependence here is given by the number
of initial binary collisions $N_{coll}$. The actual comparison in
Fig.~\ref{PsiPiRHI} indicates that the statistical model by
Andronic {\it et al.} \cite{PBM07} predicts a sizeably larger
$J/\Psi$ to $\pi$ ratio at midrapidity for peripheral and
semi-peripheral reactions than the microscopic HSD results for the
different scenarios. For central reactions - where an approximate
equilibrium is achieved - all scenarios give roughly the same
ratio. In full $4 \pi$ phase space the HSD results indicate also a
slightly higher $J/\Psi$ to $\pi$ ratio in the 'comover' model
relative to the 'melting' scenario but both ratios only weakly
depend on centrality roughly in line with the statistical model of
Gorenstein and Gazdzicki ~\cite{MG**2} (orange straight line).
Consequently, only peripheral reactions of heavy nuclei might be
used to disentangle the different scenarios at top SPS energies at
midrapidity (or in full phase space).

The situation is different for Au+Au collisions at the top RHIC
energy as may be extracted from Fig. \ref{PsiPiRHI} where the
$J/\Psi$ to pion ratio (l.h.s.: at midrapidity; r.h.s.: for $4\pi$
acceptance) is shown again as a function of $N_{coll}$. The
standard 'comover' model (dashed blue lines) is only shown for
reference but is unrealistic according to the analysis in Section
6. We find that the 'comover' model with early pre-hadronic charm
interactions (solid red line with stars, l.h.s.) is very close to
the statistical hadronization model \cite{PBM07} (solid black
line) at midrapidity except for very peripheral collisions. The
'threshold melting' scenario follows the trend in centrality but
is down by about 30\%. Thus at midrapidity there is no essential
extra potential in differentiating the scenarios. Considering the
full $4 \pi$ acceptance (r.h.s.) we find a practically constant
$J/\Psi$ to pion ratio for $N_{coll} > $ 200 from the HSD
calculations as expected from the statistical model, however, the
early model of Gorenstein and Gazdzicki ~\cite{MG**2} is down by
about a factor of $\sim$ 10 (and may be ruled out by present
data).

\subsection{Excitation functions}

\begin{figure}
%\phantom{a}
%\vspace*{-0.5cm}
\centerline{\psfig{figure=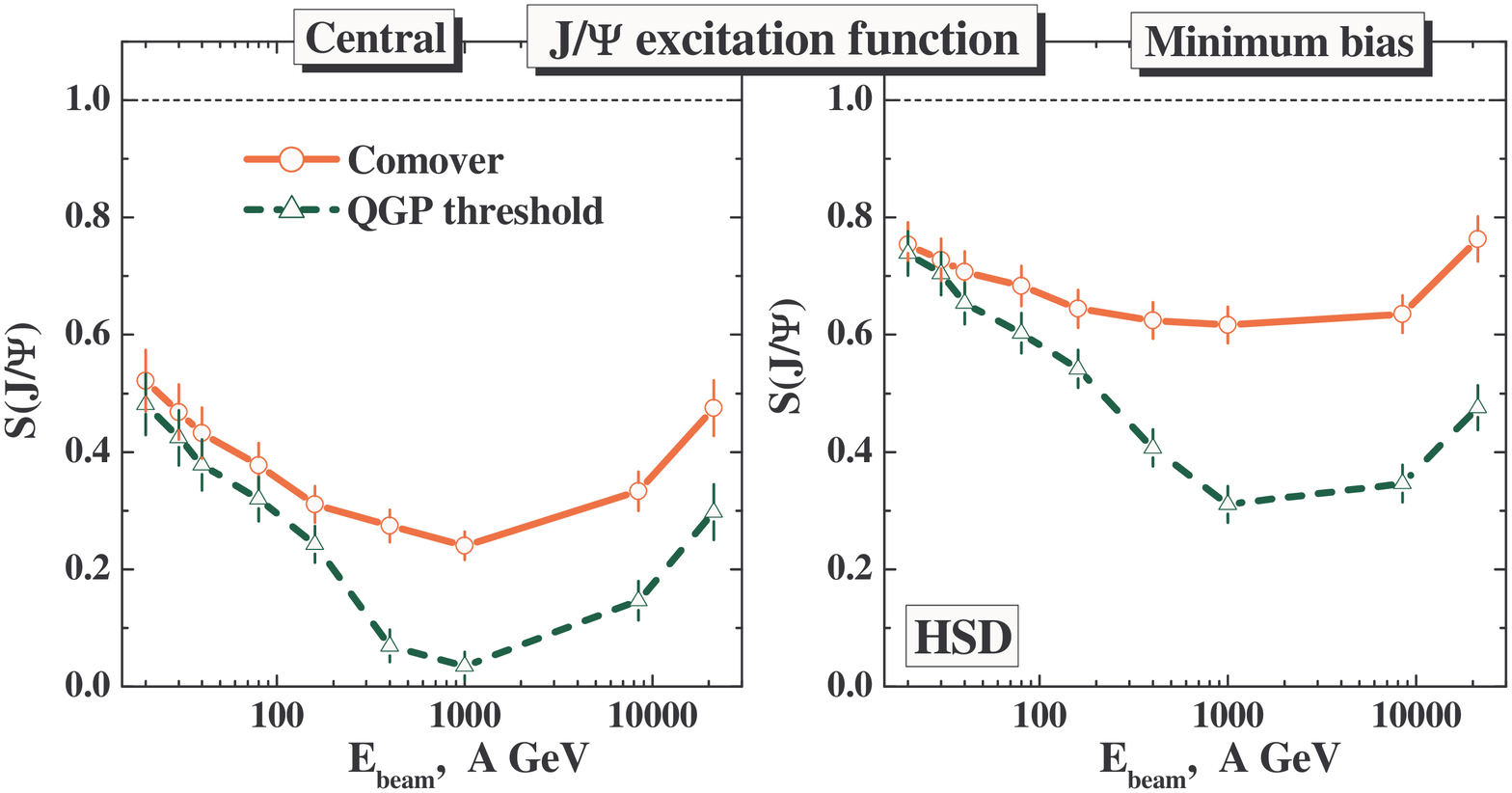,width=\textwidth}}
\centerline{\psfig{figure=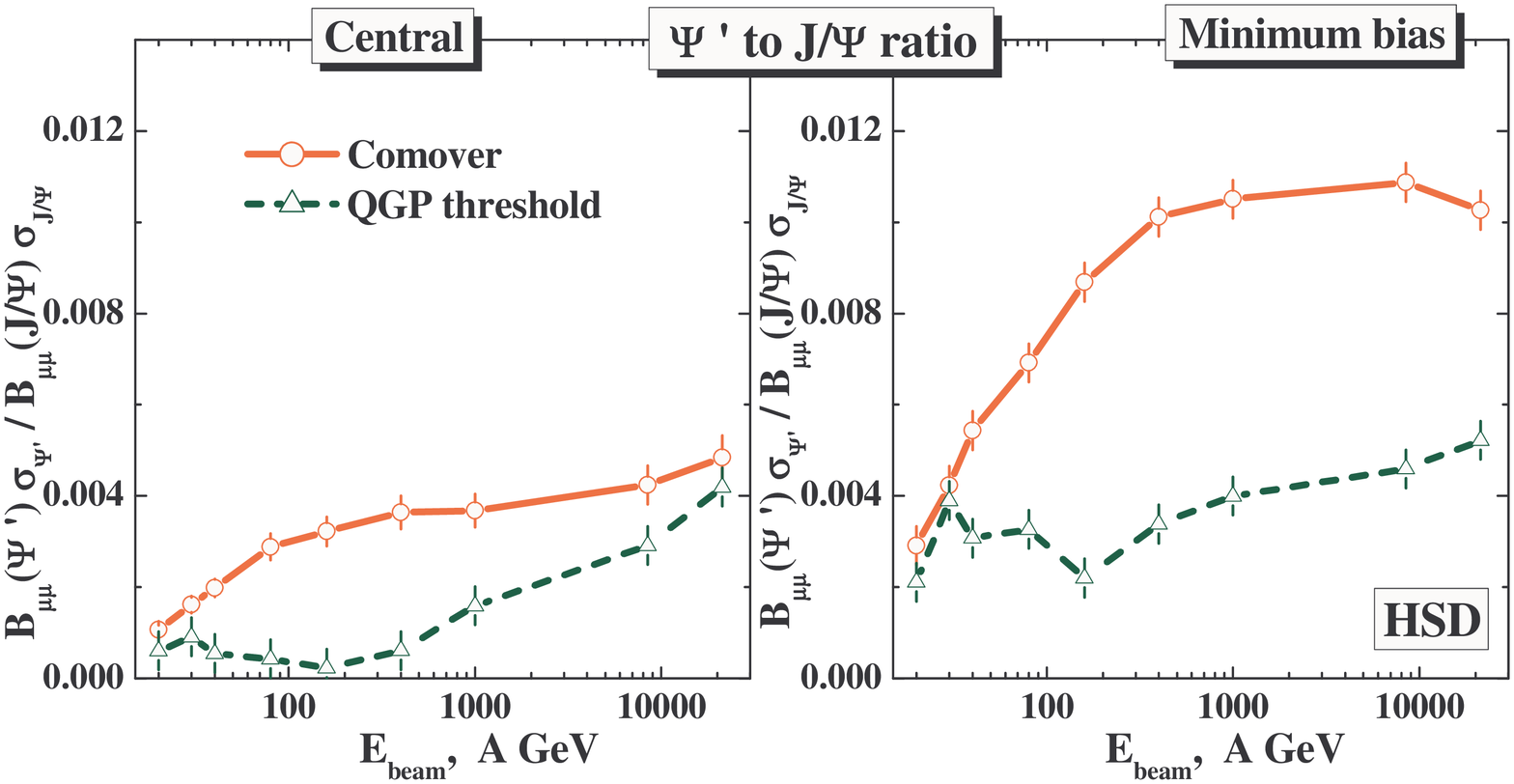,width=\textwidth}}
\caption{upper part: The excitation function for the $J/\Psi$
survival probability in the `QGP threshold melting + hadronic
recombination' scenario (dashed green lines with triangles) and the
`comover absorption + recombination' model (solid red lines with
circles) for central (l.h.s.) and minimum bias Au+Au reactions
(r.h.s.) as a function of the beam energy. Lower part: The
$\Psi^\prime$ to $J/\Psi$ ratio for the same reactions as in
the upper part of the figure in the `QGP threshold melting +
hadronic recombination' scenario (dashed green lines with triangles) and
the `comover absorption + recombination' model (upper solid red lines with
circles ).} \label{JPsiEx}
\end{figure}

In this Subsection we present the excitation functions for the $J/\Psi$
survival probability in Au + Au collisions from FAIR to top RHIC
energies in the different scenarios in order to allow for a further
distinction between the different concepts. The results of our HSD
calculations are presented in the upper part of Fig. \ref{JPsiEx} for
the `QGP threshold melting + hadronic recombination' scenario
(dashed green lines with open triangles) and the `comover absorption + recombination'
model (solid red lines with open circles) for central (l.h.s.) and minimum bias
(r.h.s.) Au+Au reactions  as a function of the beam energy. We find
that from FAIR energies of 20 - 40 A$\cdot$ GeV up to top SPS energies
of 158 A$\cdot$ GeV there is no significant difference for the
$J/\Psi$ survival probability in case of central collisions. The
differences here show mainly up in the full RHIC energy range where the
`QGP threshold melting + hadronic recombination' scenario leads to a
substantially lower $J/\Psi$ survival probabilities. In case of minimum
bias collisions the `comover absorption + recombination' model (solid
lines) leads to a roughly energy independent $J/\Psi$ survival
probability whereas the `QGP threshold melting + hadronic
recombination' scenario shows lower $J/\Psi$ survival probabilities
(lower dashed green lines) for laboratory energies above $\sim$ 100
A$\cdot$GeV due to a larger initial melting of $J/\Psi$ at high energy
density.

A clearer distinction between the different concepts is offered by the
excitation functions for the $\Psi^\prime$ to $J/\Psi$ ratio in Au + Au
collisions. The calculated results are shown in the lower part of Fig.
\ref{JPsiEx} for the `QGP threshold melting + hadronic recombination'
scenario (dashed green lines with open triangles) and the `comover absorption +
recombination' model (solid red lines) for central (l.h.s.) and
minimum bias reactions (r.h.s.). Here the $\Psi^\prime$ is already
melting away in central Au+Au reactions in the `QGP threshold melting'
scenario at bombarding energies above 40 A$\cdot$GeV whereas a
substantial amount of $\Psi^\prime$ survives in the `comover absorption
+ recombination' model. Thus measurements of $\Psi^\prime$ suppression
at the lower SPS or top FAIR energies will clearly distinguish between
the different model concepts.

%%%%%%%%%%%%%%%%%%%%%%%%%%%%%%%%%%%%%%%%%%%%%%%%%%%%%%%%%%%%

\subsection{Elliptic flow of charm} \label{flow}

\begin{figure}
%\phantom{a}
%\vspace*{-0.5cm}
\centerline{\psfig{figure=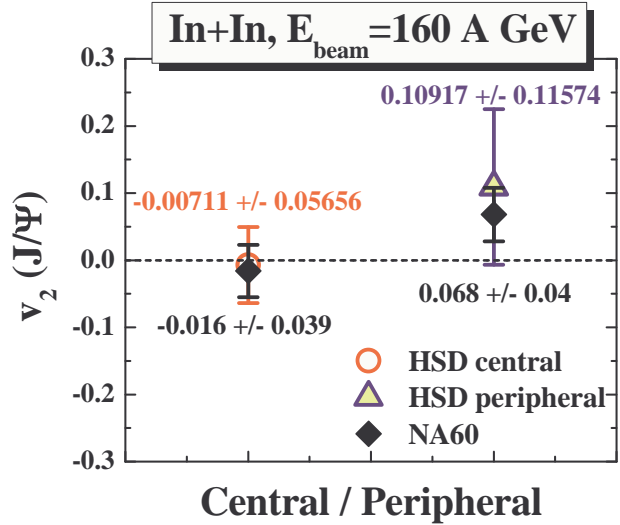,width=0.6\textwidth}}
\caption{Elliptic flow $v_2$ of $J/\Psi$'s produced in central and
peripheral  $In+In$ collisions at 158~A$\cdot$GeV beam energy in
the hadronic `comover' mode of HSD (open circle and open triangle)
compared to the NA60 data~\cite{NA602007} represented by black
diamonds.} \label{NA60v2}
\end{figure}

The elliptic flow of particles defined as
\begin{equation}
\label{v2} v_2(y,p_T) =
\left\langle\frac{p_x^2-p_y^2}{p_T^2}\right\rangle_{y,p_T}
\end{equation} (with $p_T^2= p_x^2 + p_y^2$)
provides additional information on the collective currents and pressure
evolution in the early phase of the complex reaction \cite{StoeckerV12}
 since it is driven by different pressure gradients in case of
nonvanishing spatial anisotropy $\epsilon_2 = < \frac{y^2 - x^2}{y^2 +
x^2}>$. Since $\epsilon_2$ decreases fast during the expansion of a
noncentral reaction the magnitude of $v_2$ gives information about the
interaction strength or interaction rate of the early medium.

In Fig.~\ref{NA60v2} we test the HSD result for $v_2 (J/\Psi)$ at
SPS in the purely hadronic 'comover' scenario in comparison to the
data for $v_2$ of the NA60 collaboration for In+In collisions
~\cite{NA602007}. In central collisions the elliptic flow is
practically zero both in the calculation as well as in the
experiment whereas in peripheral reactions a nonzero flow emerges.
The agreement (within error bars) between the theory and the data
indicates that in line with the reproduction of the $J/\Psi$
suppression data \cite{Olena} the low amount of $v_2$ does not
point towards additional strong partonic interactions.
Consequently, the present measurements of $J/\Psi$ elliptic flow
at SPS energies do not provide further constraints on the model
assumptions.

\begin{figure}
%\phantom{a}
%\vspace*{-0.5cm}
\centerline{\psfig{figure=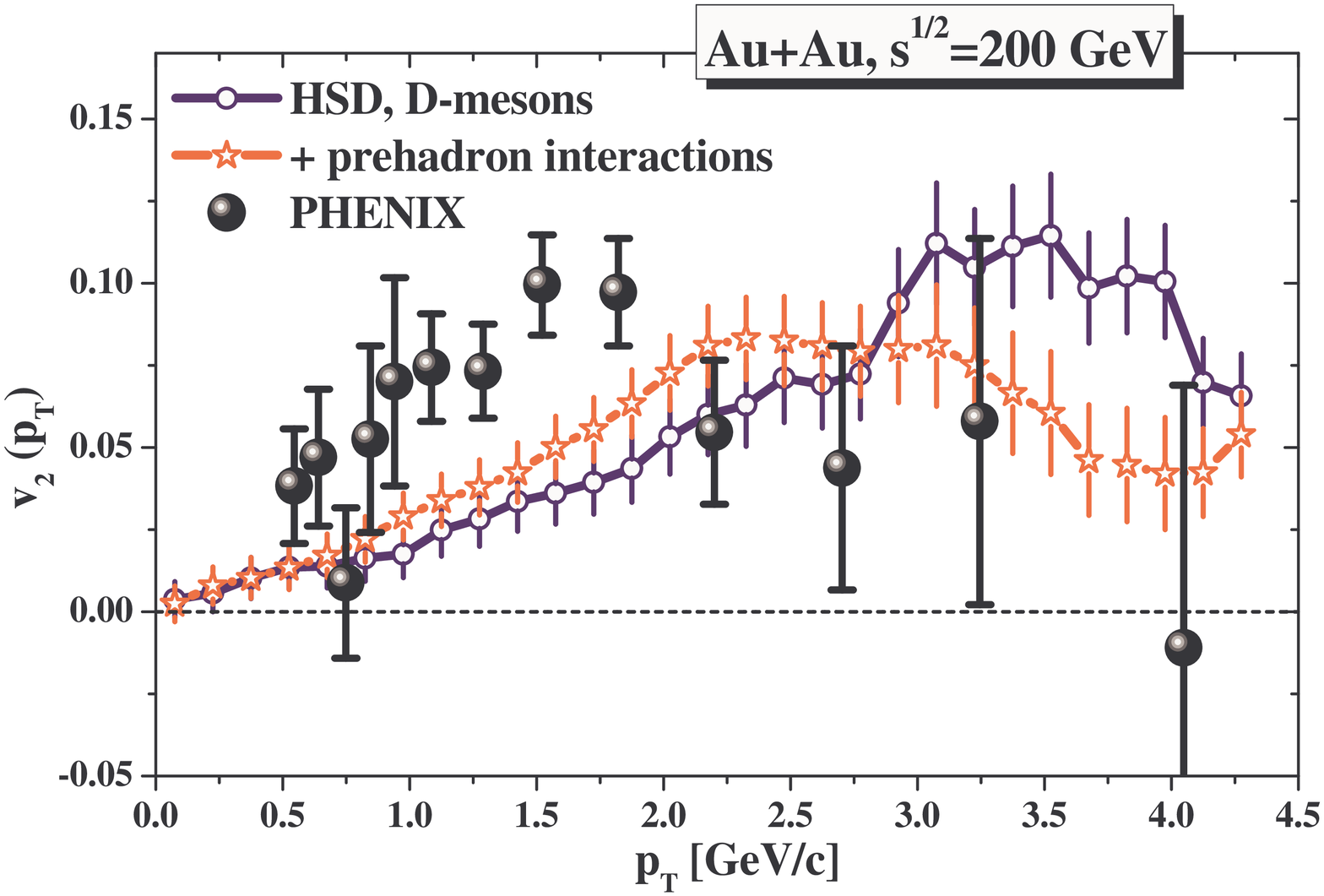,width=0.8\textwidth}}
\caption{Elliptic flow of $D$-mesons produced in $Au+Au$
collisions at $\sqrt{s}=200$~GeV as a function of $p_T$ from HSD
(solid blue line with open circles) in
comparison to the PHENIX data~\cite{PHENIXv2D} on $v_2$ of
non-photonic electrons. The red line with open stars shows the HSD
result for the $v_2$ of $D$-mesons when including additionally
pre-hadronic charm interactions as described in Section 6.}
\label{Dv2}
\end{figure}

The situation, however, is different for the collective flow of
$D$-mesons at top RHIC energies. In Fig. \ref{Dv2} we show the elliptic
flow of $D$-mesons produced in $Au+Au$ collisions at $\sqrt{s}=200$~GeV
as a function of the transverse momentum $p_T$ in HSD (solid blue line
with open circles) compared to the PHENIX
data~\cite{PHENIXv2D} on $v_2$ of non-photonic electrons. Here the
elliptic flow of $D$-mesons is clearly underestimated in the standard
HSD model (cf. Ref.  \cite{brat05}). Only when including pre-hadronic
charm interactions - as described in Section 6 - the elliptic flow
increases (red line with open stars) but still stays clearly below
the PHENIX data for $p_T < $ 2 GeV/c. We thus have to conclude that the
modeling of charm interactions by pre-hadronic interactions - as
described in Section 6 - does not provide enough interaction strength
in the early phase of the collision. Quite remarkably this finding is
again fully in line with the underestimation of high $p_T$ hadron
suppression \cite{HPT1} as well as far-side jet suppression \cite{HPT2}
in the pre-hadronic interaction model. Independently, also the charm
collective flow points towards strong partonic interactions in the
early reaction phase beyond the pre-hadronic scattering incorporated so
far.

Since a large fraction of $J/\Psi$'s in central Au+Au collisions at
RHIC are created by $D- {\bar D}$ recombination, the elliptic flow of
$J/\Psi$'s  obtained from HSD in the comover (purely hadronic) case is
comparatively small, too, and should not be in accord with future
experimental data. We consequently discard an explicit representation
of the $J/\Psi$ elliptic flow at RHIC energies since the calculations
show the $v_2$ of charmonium to be very close to the $D$-meson flow
within error bars.

%**************************************************************************

\section{Summary}
Our present study essentially  completes the investigations of
charm production, propagation and chemical reactions within the
HSD transport approach initiated more than a decade ago
\cite{Cass97,CassKo}. The present systematic investigation extends
earlier work to RHIC energies and clearly shows - as advocated
before \cite{Olena2} - that the traditional concepts of
`charmonium melting' in a QGP state as well as the hadronic
`comover absorption and recreation model' are in severe conflict
with the data from the PHENIX Collaboration at RHIC energies
whereas both model assumptions work reasonably well at top SPS
energies \cite{Olena}.

The essential new result of this work is that (at top RHIC
energies) we find evidence for strong interactions of charm with
the pre-hadronic medium  from comparison to recent data from the
PHENIX Collaboration \cite{PHENIXNov06}. In particular,
pre-hadronic interactions (of unformed hadrons) with charm lead to
dramatically different rapidity distributions for $J/\Psi$'s and
consequently to a substantially modified ratio $R_{AA} ^{forward}
(J/\Psi)$ to $R_{AA} ^{mid} (J/\Psi)$ compared to earlier
calculations/predictions.

Further results of the present microscopic transport
study may be stated as follows:
\begin{itemize}
\item{The $J/\Psi$ suppression in $d+Au$ collisions at $\sqrt{s}$ =
200 GeV is only roughly compatible with the charmonium absorption
on nuclei as observed at SPS energies in $p+A$ reactions. We find
a clear indication for shadowing effects at forward rapidity, but
a conclusive answer about the size of this effect is not possible
due to the statistical error bars in both the experimental data
and the calculations. A proper answer can only be given by future
high statistics data that allow to fix the scale of shadowing in a
model independent way.}
\item{The $\Psi^\prime$ to $J/\Psi$ ratio is found to be crucial in disentangling
the different charmonium absorption scenarios. This result
essentially emerges from the early dissociation of $\Psi^\prime$
above the critical energy density $\epsilon_c \approx $ 1
GeV/fm$^3$ in the `QGP melting scenario' whereas the $\Psi^\prime$
in the `comover model' survives to higher energy densities.}
\item{A comparison of the transport calculations to the statistical
model of Gorenstein and Gazdzicki \cite{MG**2} (in $4 \pi$ acceptance) or the
statistical hadronization model of Andronic  {\it et al.} \cite{PBM07} (at
midrapidity)  shows differences in the energy
as well as centrality dependence of the $J/\Psi$ to pion ratio,
which might be exploited experimentally to discriminate the
different concepts. }
\item{The collective flow of charm in the HSD transport appears
compatible with the data at SPS energies, but the data are
substantially underestimated  at top RHIC energies (cf. Fig. 16).
This not only holds for the standard hadronic comover scenario,
but also when including interactions of charm with pre-hadronic
states (unformed hadrons). Consequently the large elliptic flow
$v_2$ of charm seen experimentally has to be attributed to early
interactions of non-hadronic degrees of freedom.}
\end{itemize}

The open problem - and future challenge - is to incorporate
explicit partonic degrees of freedom in the description of
relativistic nucleus-nucleus collisions and their transition to
hadronic states in a microscopic transport approach. On the
experimental side, further differential spectra of charmonia and
open charm mesons then will constrain the transport properties of
charm in the early non-hadronic phase of nucleus-nucleus
collisions at RHIC (and possibly at SPS or even FAIR energies).

%------------------------------------------------------------------------
\section*{Acknowledgements}

We acknowledge stimulating correspondence with A.~Andronic,
T.~Gunji, D.~Kim and J.~Skullerud as well as helpful discussions
with P.~Braun-Munzinger, M.~Gorenstein, R.~Granier de Cassagnac,
K.~Redlich, J.~Stachel and H. St\"ocker. Furthermore, O. L. and
E.L.B. would like to thank the BMBF for financial support.

%\bibliographystyle{h-physrev3}
%\bibliography{HSDcharm}

\end{document}